\documentclass{JFM-FLM_Au}
\makeatletter
\def\oddabsfooterflag{\hbox to \hsize{\hfil{\cppagefont\thepage}}}
\def\evenabsfooterflag{\hbox to \hsize{{\cppagefont\thepage}\hfil}}
\makeatother
\usepackage{fix-cm}
\usepackage{amsmath}
\usepackage{graphicx}
\usepackage{rotating} 
\usepackage{siunitx}
\usepackage{booktabs}
\usepackage{bm} 
\usepackage{hyperref}
\usepackage[table,xcdraw]{xcolor}
\usepackage[justification=justified]{caption}
\usepackage{natbib}
\usepackage{enumitem} 
\DeclareMathAlphabet{\mathsf}{OT1}{cmss}{m}{n} 
\hypersetup{
  colorlinks = true,
  urlcolor  = blue,
  citecolor = black,
}

\newcommand{\Rep}{\mathrm{Re}_p}

\newcommand{\dd}{\mathrm{d}}

\renewcommand{\p}{\mathbf{p}}
\newcommand{\ghat}{\hat{\mathbf{g}}}
\newcommand{\w}{\mathbf{w}}
\newcommand{\I}{\mathbf{I}}
\newcommand{\E}{\mathbf{E}}
\newcommand{\W}{\mathbf{W}}
\newcommand{\Sv}{\mathrm{Sv}}

\title{Orientation and slip statistics of spheroids settling in turbulence}

\lefttitle{A. Roy}
\righttitle{Orientation and slip statistics of settling spheroids}

\author{Anubhab Roy\aff{1}}

\affiliation{\aff{1}Department of Applied Mechanics and Biomedical Engineering,
Indian Institute of Technology Madras, Chennai 600036, India}

\corresau{Anubhab Roy, anubhab@iitm.ac.in}

\begin{document}

\maketitle

\begin{abstract}
Small spheroids settling through turbulence are turned broadside-on by fluid inertia and tilted away by velocity gradients. We describe their orientation and slip statistics through the crossover from random orientation to alignment, in the overdamped limit. With quasi-static Stokes drag the leading inertial torque is an exact gradient flow on the orientation sphere, its potential quadratic in the cosine of the tilt and only rescaled by mobility anisotropy. The orientation performs an overdamped random walk in a quadratic well, and isotropic Gaussian white noise gives exactly a Boltzmann distribution. For forcing of finite correlation time, a closure keeps this form with its width set by linear response; it recovers the uniform density as settling vanishes and, for fixed-orientation forcing statistics, the aligned coefficient of rapid-settling theory, and errs most in the crossover. The root-mean-square lateral slip peaks where the alignment rate approaches the Kolmogorov rate, as faster settling strengthens the sideways drift at a given tilt but suppresses the tilt; the nearly shape-independent ratio, which tensor-forced simulations reproduce, lets a measured peak constrain the torque coefficient or dissipation rate for given forcing statistics. Dissipation intermittency enters as a lognormal superposition of Boltzmann distributions, which fattens the tails but leaves the leading aligned mean-square tilt unchanged. The superposition reproduces the shape of plate orientation distributions from direct numerical simulations, where a single Boltzmann distribution fails; published moments confirm the predicted spread beyond the weakest torques, and at strong alignment the measured fourth-order tilt moment lies between the two predictions.
\end{abstract}

\section{Introduction}

The orientation of small nonspherical particles settling through a turbulent fluid controls their optical signatures, their settling speeds and their collision rates. Ice crystals falling broadside-on produce the specular glints, polarimetric signals and enhanced backscatter used in lidar and radar retrievals \citep{Pruppacher1997,Breon2004,Yang2003}, which return orientation averages, and observations place real cirrus between the two limits usually assumed, random orientation and perfect alignment \citep{Breon2004,Sassen2001,Noel2010}. The width of the orientation density matters as much as its mean: for plates above $200$\,\si{\micro\metre}, tilt widths across the plausible range $0.5^\circ$--$5^\circ$ change computed lidar backscatter by more than two orders of magnitude at fixed size \citep{Okamoto2020}. The same statistics enter collision models for ice crystals \citep{Naso2018,Jucha2018,Sheikh2022,Sheikh2024} and set the orientation of frazil discs in supercooled water \citep{Daly1984}.

The torque due to fluid inertia is well understood in the near-viscous regime. At small but finite particle Reynolds number $\Rep$ a fluid-inertial torque turns a settling spheroid broadside-on to its slip. It was obtained for slender bodies by \cite{Khayat1989} and for arbitrary aspect ratio by \cite{Dabade2015}, and an inertial slender-body theory extends it to diameter-based Reynolds numbers of $O(10)$ \citep{Joshi2026}. Experiments confirm the alignment in quiescent fluid \citep{Lopez2017,Roy2019} and show that in air, where the particle is far denser than the fluid, the approach to it oscillates and the orientation fluctuates strongly \citep{Bhowmick2024}.

The fluctuations about the broadside orientation are less well understood. \cite{Klett1995}, following \cite{Cho1981}, balanced a restoring torque against an assumed white-noise turbulent torque; his density has the form derived below, but with coefficients assumed rather than obtained from the equations of motion. Rapid-settling theories obtain those coefficients for arbitrary aspect ratio \citep{Menon2017,Gustavsson2017,Gustavsson2019,Anand2020,Roy2023}, \cite{Gustavsson2021} add particle inertia, \cite{Sheikh2020} estimate when the fluid-inertial torque outweighs the turbulent one, and \cite{Roy2023} map a shear-rate distribution onto an orientation density in that limit; direct numerical simulations span the transition from disorder to alignment \citep{Siewert2014,Jucha2018,Anand2020,Pumir2026}. The derived results are, however, small-tilt expansions or rapid-settling asymptotics, whereas for crystals of tens of microns in cloud turbulence the alignment parameter is of order unity \citep{Gustavsson2021,Pumir2026}: the crossover, where those expansions fail and the available results come from simulation and from experiments on finite-size particles \citep{Roy2023}.

We treat that crossover in the overdamped limit, in which the response times of the particle are short and its inertia does not enter (Sec.~\ref{sec:application} sets out its range of validity). The inertial torque is an exact gradient flow on the sphere of orientations, with a potential quadratic in the cosine of the tilt, and the Jeffery forcing at a fixed orientation has no preferred direction in the tangent plane. For white-noise forcing the stationary density is then a Boltzmann distribution, as for orientable Brownian particles \citep{Brenner1974}, here exponential in the squared cosine of the tilt. Turbulent forcing has a finite correlation time, which leaves the orientation alone non-Markovian. We keep the Boltzmann form, fix its concentration from the exact linear response, and test that closure against stochastic simulations; it returns the uniform density as settling vanishes and, with fixed-orientation forcing statistics, the rapid-settling coefficient when the torque is strong (Sec.~\ref{sec:correspondence}).

\begin{figure}
\includegraphics[width=\textwidth]{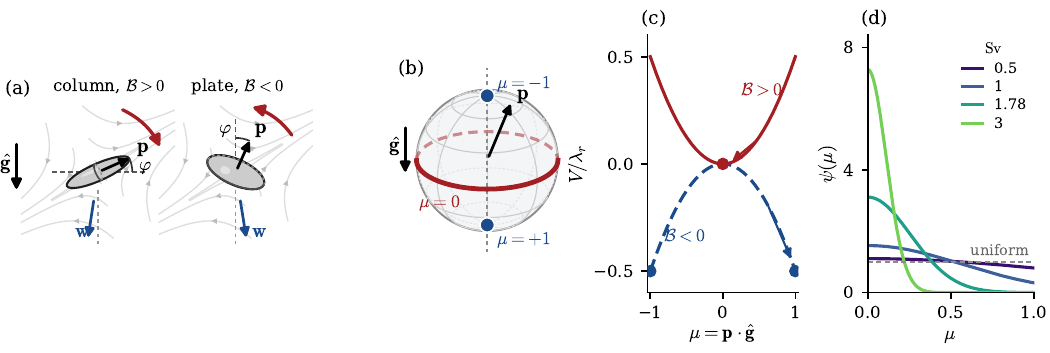}
\caption{Schematic of the problem. (a) A spheroid settling with slip $\mathbf w$ through a locally linear flow (dashed lines, the vertical and the horizontal): the anisotropic mobility turns $\mathbf w$ away from $\ghat$, Eq.~\eqref{eq:slipvec}, and the fluid-inertial torque (red arcs) reduces the tilt $\varphi$. (b) Attracting sets: the girdle $\mu=0$ for columns ($\mathcal{B}>0$) and the poles $\mu=\pm1$ for plates ($\mathcal{B}<0$). (c) The orientational potential \eqref{eq:lamr}. (d) The density of the $\kappa=10$ column at four settling numbers, at $F=0.316$.}
\label{fig:schematic}
\end{figure}

The even moments of that density give in closed form the orientation averages a scattering calculation needs and the slip statistics. The lateral slip first grows and then decays as the particle settles faster, peaking where the alignment rate is close to the Kolmogorov rate, at a ratio of the two that depends only weakly on shape, so the peak constrains the torque coefficient or, with the torque known, the dissipation rate, given the forcing statistics. The structure carries over to orientation-dependent torques and to dissipation intermittency, which enters as a lognormal superposition of Boltzmann densities and fits orientation densities of settling plates from direct numerical simulation much better than a single one. The published moments of those simulations follow the predicted spread except at the weakest torques, while at strong torque their fourth-order tilt moment lies between the predictions of the single density and the superposition. Figure~\ref{fig:schematic} sets out the problem and the notation.

Sections~\ref{sec:orient}--\ref{sec:stationary} set out the orientation dynamics, the turbulent forcing and the stationary statistics, Secs.~\ref{sec:slip} and \ref{sec:peak} the slip and its maximum, Sec.~\ref{sec:beyond} the extensions, Sec.~\ref{sec:tests} the tests against simulated turbulence, and Sec.~\ref{sec:application} the application to atmospheric ice and frazil. Appendix~\ref{app:validation} lists the checks behind the closed-form results.

\section{Orientation dynamics}
\label{sec:orient}

\subsection{Equations of motion}

Consider a rigid spheroid with symmetry semi-axis $c$, transverse semi-axis $a$, aspect ratio $\kappa=c/a$ and unit symmetry axis $\p(t)$, settling through homogeneous isotropic turbulence. A single $\kappa\in(0,\infty)$ covers elongated and flattened shapes, which enter the orientation dynamics through the Bretherton constant
\begin{equation}
\mathcal{B}=\frac{\kappa^2-1}{\kappa^2+1}\in(-1,1)
\label{eq:bretherton}
\end{equation}
and, through the mobility, the settling factors $(\hat U_\parallel,\hat U_\perp)$ of Sec.~\ref{sec:renorm}. The sign of $\mathcal{B}$ separates prolate columns ($\mathcal{B}>0$) from oblate plates ($\mathcal{B}<0$) and, as Sec.~\ref{sec:topology} shows, decides whether the particle is attracted to the girdle or to the poles. The particle is smaller than the Kolmogorov length $\eta$, so it samples a locally linear flow; its translational and rotational response times are short, so the slip velocity $\w$ relative to the local fluid is set quasi-statically by gravity and Stokes drag and its rotation is overdamped; and its slip Reynolds number is small. We keep Jeffery's rotation and the leading inertial torque due to the slip, and neglect the inertial corrections due to the ambient velocity gradient. The orientation then evolves as
\begin{equation}
\dot\p \;=\; \W\cdot\p \;+\; \mathcal{B}\,(\I-\p\p)\cdot\E\cdot\p \;+\; \dot\p_I ,
\label{eq:ndot}
\end{equation}
the first two terms being Jeffery's rotation \citep{Jeffery1922} in the spin $\W$ and strain $\E$ of the local flow, and the last the fluid-inertial contribution. At $O(\Rep)$ the inertial torque on a spheroid in a uniform slip $\w$ has the exact vector structure $\propto(\w\cdot\p)(\w\times\p)$ \citep{Khayat1989,Dabade2015,Jiang2021}, so that, after division by the transverse rotational resistance,
\begin{equation}
\dot\p_I \;=\; -\,s_\kappa\,\frac{F_0}{\nu}\,(\w\cdot\p)\,(\I-\p\p)\cdot\w ,\qquad s_\kappa=\operatorname{sgn}\mathcal{B},
\label{eq:torque}
\end{equation}
where $F_0(\kappa)>0$ for $\kappa\neq1$ is the bare dimensionless torque coefficient (Appendix~\ref{app:mobility}) and $s_\kappa$ carries the prolate--oblate distinction: the torque turns any spheroid broadside-on to its slip, with the symmetry axis across the slip for a column and along it for a plate. Below, $F$ without a subscript denotes the effective coefficient defined by Eq.~\eqref{eq:lamr}; Sec.~\ref{sec:renorm} relates it to $F_0$.

\subsection{Gradient-flow structure}
\label{sec:gradflow}

For slip locked to gravity, $\w=w\,\ghat$, Eq.~\eqref{eq:torque} reads $\dot\p_I=-s_\kappa(F_0w^2/\nu)\,\mu\,(\I-\p\p)\cdot\ghat$ with $\mu=\p\cdot\ghat$, which drives $\mu$ to $0$ for a column and to $\pm1$ for a plate. Since $(\I-\p\p)\cdot\nabla_\p V=V'(\mu)(\I-\p\p)\cdot\ghat$ for any $V(\mu)$, the drift is exactly the negative surface gradient of the potential
\begin{equation}
V(\p) \;=\; \tfrac12\,s_\kappa\,\lambda_r\,\mu^2 ,
\qquad
\lambda_r=\frac{F w^2}{\nu}>0,
\qquad
\lambda_r\tau_K = F\,\Sv^2 ,
\label{eq:lamr}
\end{equation}
where $\tau_K=(\nu/\varepsilon)^{1/2}$ is the Kolmogorov time and $\Sv=w/u_\eta$ the settling number based on the Kolmogorov velocity $u_\eta=(\nu\varepsilon)^{1/4}$; we call $\lambda_r\tau_K=F\Sv^2$ the alignment parameter. For slip locked to gravity $F=F_0$; Sec.~\ref{sec:renorm} shows that an orientation-dependent slip changes only this coefficient. More generally, any axisymmetric drift $g(\mu)(\I-\p\p)\cdot\ghat$ is a gradient flow on the sphere with $V'(\mu)=-g(\mu)$, a fact used in Sec.~\ref{sec:beyond}.

The strain term has the same structure. For any symmetric $\mathsf Q$,
\begin{equation}
\nabla_s\bigl(\tfrac12\,\p\cdot\mathsf Q\cdot\p\bigr)=(\I-\p\p)\cdot\mathsf Q\cdot\p ,
\label{eq:quadgrad}
\end{equation}
where $\nabla_s$ is the gradient on the unit sphere, the rotary-potential construction of \cite{Brenner1974}. With $\mathsf Q=\mathcal{B}\E$ and $\mathsf Q=-s_\kappa\lambda_r\ghat\ghat$, Eq.~\eqref{eq:ndot} becomes
\begin{equation}
\dot\p=\W\cdot\p+\nabla_s X,
\qquad
X=\tfrac12\,\p\cdot\bigl(\mathcal{B}\E-s_\kappa\lambda_r\ghat\ghat\bigr)\cdot\p ,
\label{eq:gradform}
\end{equation}
a rigid rotation at the ambient angular velocity together with gradient ascent on a single quadratic form. The rotation, $\W\cdot\p=-\tfrac12\,\p\times\nabla_s(\bm\omega\cdot\p)$ with $\bm\omega$ the vorticity, is not a gradient. A rotation history prescribed independently of $\p$ turns the sphere rigidly and leaves a uniform density uniform; once the slip, and with it the path, depends on $\p$, so does the vorticity history, and the random-orientation limit rests instead on the joint statistical isotropy of $\p$ and the gradients (Sec.~\ref{sec:forcing}). For a steady flow, by contrast, the rotary flux vanishes only if the flow is irrotational \citep{Brenner1974}; here the vorticity belongs to the stochastic forcing, and the potential comes from gravity.

The axisymmetric potential has two further consequences. Its drift has no azimuthal component, so the azimuth about gravity is exactly neutral, and its relaxation is solvable at any tilt. In the vertical plane that contains $\p$, which the drift leaves fixed, let $\varphi$ be the signed angle of $\p$ from the attracting direction, the horizontal for a column and the vertical for a plate, positive on a chosen side; its magnitude is the tilt, $\arcsin|\mu|$ for a column and $\arccos|\mu|$ for a plate. Then $\dot\varphi=-\tfrac12\lambda_r\sin2\varphi$, and $\tan\varphi$ decays as $e^{-\lambda_rt}$. This is the overdamped limit of a damped oscillator. With rotational inertia, and the slip still quasi-static, the same torque gives $\tau_r\ddot\varphi+\dot\varphi=-\tfrac12\lambda_r\sin2\varphi$, with $\tau_r$ the rotational response time, so that small tilts overshoot the broadside orientation when $4\lambda_r\tau_r>1$. Heavy spheroids settling in still air oscillate in this way \citep{Bhowmick2024}, though their full dynamics also couples the tilt to a lagging slip; for $\lambda_r\tau_r\ll1$, the condition of Sec.~\ref{sec:application}, the inertial term drops out and the gradient flow remains. Since the restoring drift is $-\lambda_r\cos^2\varphi\tan\varphi$, a forcing that carries the same factor $\cos^2\varphi$ keeps the problem linear in $\tan\varphi$, as for a thin disc in a vertical simple shear of rate $\gamma$, for which $\dot\varphi=-\gamma\cos^2\varphi-\tfrac12\lambda_r\sin2\varphi$ \citep{Roy2023}. In $t_\varphi=\tan\varphi$,
\begin{equation}
\dot t_\varphi=-\lambda_r t_\varphi-\gamma,
\label{eq:tanode}
\end{equation}
so that for constant $\gamma$
\begin{equation}
\tan\varphi(t)=\Bigl[\tan\varphi(0)+\frac{\gamma}{\lambda_r}\Bigr]e^{-\lambda_r t}-\frac{\gamma}{\lambda_r},
\label{eq:tanphi}
\end{equation}
which approaches the equilibrium
\begin{equation}
\tan\varphi=-\gamma/\lambda_r
\label{eq:maptan}
\end{equation}
with no small-angle assumption. For a time-dependent $\gamma(t)$ the tangent is an exact linear filter of the shear rate, Gaussian when $\gamma(t)$ is Gaussian, a property special to this planar problem. \cite{Roy2023} map a shear-rate distribution through Eq.~\eqref{eq:maptan} to obtain an orientation density (Sec.~\ref{sec:correspondence}); the full competition between the shear and settling torques is mapped in \cite{MishraShear2026}.

\subsection{Attracting sets}
\label{sec:topology}

Columns and plates both settle broadside-on, but their attracting sets differ (figure~\ref{fig:schematic}b). A column lies with its long axis horizontal: a tilt out of the horizontal is restored, while a swing within the horizontal plane meets no resistance, every horizontal direction being equally broadside-on. A plate settles face-down, and a tilt in any direction meets a restoring torque. In terms of the potential, the minimum of $V$ for $\mathcal{B}>0$ is the equatorial girdle $\mu=0$, with one restored and one neutral tangent direction; for $\mathcal{B}<0$ it is the pole pair $\mu=\pm1$ ($\mu=+1$ the lower pole, with $\ghat$ pointing downwards), a single physical orientation by fore--aft symmetry, with both tangent directions restored at the same rate $\lambda_r$. With the two components equally forced (Sec.~\ref{sec:forcing}), the leading mean-square tilt of a plate is therefore twice that of a column at equal concentration. The same factor of two appears in the persistent limit of \cite{Anand2020} and, as a shape coefficient of one for columns and two for plates, through the inertial regimes of \cite{Gustavsson2021}. The random-orientation limits are not in the ratio two, because the spherical measure differs near a pole and near a girdle: they are $\pi^2/4-2$ for a column and $\pi-2$ for a plate [Eq.~\eqref{eq:randpro}], in the ratio $2.442$.

\subsection{Effect of mobility anisotropy}
\label{sec:renorm}

At small translational Stokes number the quasi-static balance gives $\w=W_{\rm eq}\,\hat{\mathsf M}(\p)\cdot\ghat$, where $W_{\rm eq}$ is the Stokes settling speed of the volume-equivalent sphere, $\Sv_{\rm eq}=W_{\rm eq}/u_\eta$, and $\hat{\mathsf M}=\hat U_\parallel\p\p+\hat U_\perp(\I-\p\p)$ is the dimensionless mobility, with $\hat U_\parallel(\kappa)$ and $\hat U_\perp(\kappa)$ the axial and transverse settling factors (Appendix~\ref{app:mobility}). Thus
\begin{equation}
\w \;=\; W_{\rm eq}\bigl[\hat U_\perp\,\ghat + g_M\,\mu\,\p\bigr],
\qquad g_M\equiv\hat U_\parallel-\hat U_\perp ,
\label{eq:slipvec}
\end{equation}
so both the direction and the magnitude of the slip depend on orientation, with $|\w|$ running between $\hat U_\parallel W_{\rm eq}$ and $\hat U_\perp W_{\rm eq}$. Since the torque \eqref{eq:torque} is quadratic in $\w$, one might expect quartic and higher powers of $\mu$ in the potential, and hence in the exponent of the density. Substituting Eq.~\eqref{eq:slipvec} and using $\hat U_\perp+g_M=\hat U_\parallel$ rules this out:
\begin{equation}
(\w\cdot\p)(\I-\p\p)\cdot\w
= W_{\rm eq}^2\,\hat U_\parallel\hat U_\perp\;\mu\,(\I-\p\p)\cdot\ghat,
\label{eq:identity}
\end{equation}
exactly linear in $\mu$. For the leading slip-induced torque with quasi-static Stokes drag, the potential therefore stays exactly quadratic, and the mobility only rescales the relaxation rate,
\begin{equation}
\lambda_r\tau_K \;=\; F_0\,\hat U_\parallel\hat U_\perp\,\Sv_{\rm eq}^2
\;=\;
\begin{cases}
F_0\,(\hat U_\parallel/\hat U_\perp)\,\Sv_b^2 & \text{(prolate)},\\[2pt]
F_0\,(\hat U_\perp/\hat U_\parallel)\,\Sv_b^2 & \text{(oblate)},
\end{cases}
\label{eq:renorm}
\end{equation}
where the torque \eqref{eq:torque} is evaluated with the instantaneous slip \eqref{eq:slipvec} and $\Sv_b$ is the settling number based on the broadside terminal speed ($W_b=\hat U_\perp W_{\rm eq}$ prolate, $\hat U_\parallel W_{\rm eq}$ oblate). Writing $R$ for the mobility ratio that multiplies $F_0$ in the broadside form,
\begin{equation}
\lambda_r\tau_K=F_{\rm eq}\,\Sv_{\rm eq}^2=F_b\,\Sv_b^2,\qquad F_{\rm eq}=F_0\hat U_\parallel\hat U_\perp,\qquad F_b=F_0R .
\label{eq:conventions}
\end{equation}
The effective $F$ of Eq.~\eqref{eq:lamr} is $F_{\rm eq}$ or $F_b$ according to which speed defines $\Sv$. Orientation statistics depend on $F\Sv^2$ alone and are the same in either convention; the slip statistics of Sec.~\ref{sec:slip} are written with $\Sv=\Sv_{\rm eq}$, and the convention matters whenever $F$ is compared with a quiescent-flow coefficient or read off an experiment. Departures from the quadratic potential need physics beyond the quasi-static Stokes mobility, such as nonlinear drag at finite $\Rep$ or unsteady slip; preferential sampling at finite Stokes number changes the forcing statistics rather than the potential.

\section{Turbulent forcing}
\label{sec:forcing}

The Jeffery terms in Eq.~\eqref{eq:ndot} are linear in the velocity gradient and contribute, at orientation $\p$, the tangential angular velocity
\begin{equation}
\mathbf{f} \;=\; \W\cdot\p + \mathcal{B}\,(\I-\p\p)\cdot\E\cdot\p ,
\label{eq:fdef}
\end{equation}
tangential because $\p\cdot\W\cdot\p=0$. Treating $\mathbf f$ as a prescribed stochastic forcing means evaluating its statistics at an externally fixed orientation; applying them along a trajectory assumes in addition that $\p$ is independent of the instantaneous velocity gradient. We derive the covariance first and then examine that assumption.

The one-time statistics follow from the isotropic covariance of the velocity gradient $A_{ij}=\partial u_i/\partial x_j$ in incompressible turbulence,
\begin{equation}
\langle A_{ij}A_{kl}\rangle
= \frac{\varepsilon}{30\nu}\bigl(4\delta_{ik}\delta_{jl}-\delta_{ij}\delta_{kl}-\delta_{il}\delta_{jk}\bigr),
\label{eq:Acov}
\end{equation}
whose symmetric and antisymmetric parts give
\begin{align}
\langle E_{ij}E_{kl}\rangle &= \frac{\varepsilon}{20\nu}\bigl[(\delta_{ik}\delta_{jl}+\delta_{il}\delta_{jk})-\tfrac23\delta_{ij}\delta_{kl}\bigr],
\label{eq:Ecov}\\
\langle W_{ij}W_{kl}\rangle &= \frac{\varepsilon}{12\nu}\,(\delta_{ik}\delta_{jl}-\delta_{il}\delta_{jk}),
\label{eq:Wcov}
\end{align}
each with trace $\varepsilon/2\nu$; the cross-covariance $\langle W_{ij}E_{kl}\rangle$ vanishes, no isotropic rank-four tensor being antisymmetric in one index pair and symmetric in the other. At fixed $\p$ the gradient statistics are invariant under rotations about $\p$, and the only tangent-plane tensor with that invariance is $\delta_{ij}-p_ip_j$, so no moment of $\mathbf f$ can single out a direction in the tangent plane and the contraction fixes only the coefficient:
\begin{equation}
\langle f_i f_j\rangle
= \sigma_f^2\,(\delta_{ij}-p_ip_j),
\qquad
\sigma_f^2=\frac{5+3\mathcal{B}^2}{60}\,\frac{\varepsilon}{\nu}
=\frac{5+3\mathcal{B}^2}{60\,\tau_K^2},
\label{eq:fvar}
\end{equation}
with total variance $2\sigma_f^2$, five parts from the vorticity and $3\mathcal{B}^2$ from the strain.

The independence assumption fails in general, because Jeffery rotation correlates $\p$ with the flow: rods align with the vorticity more than with the principal strain axis \citep{Pumir2011}, both aligning with the Lagrangian stretching direction \citep{Ni2014}. Either alignment weakens the forcing, since $\W\cdot\p$ vanishes for $\p$ along $\bm\omega$ and the projected strain for $\p$ along a strain eigenvector, so Eq.~\eqref{eq:fvar} overstates the forcing felt along a path, by more for slender columns. At the weak-torque end, with $\Sv$ small, gravity drops out and the joint statistics of $\p$ and the gradients are isotropic, so the stationary density is uniform whatever the alignment of $\p$ with the local gradient. At the strong-torque end the particle falls a distance $w/\lambda_r=\eta/(F\Sv)$ per relaxation time, less than the gradient correlation length $\ell\approx12$--$20\,\eta$ (Appendix~\ref{app:tauc}), so a realignment sees a single frozen gradient, at an orientation imposed by gravity rather than by the flow. That suffices for a plate, restored in both tangent directions, but not for a column, whose azimuth in the girdle is set by the Jeffery dynamics at any settling number. The assumption is therefore controlled at both ends for plates, at the random end only for columns, and nowhere in the crossover, where its error adds to that of the closure (Sec.~\ref{sec:conditioning}).

We model $\mathbf f$, whose time dependence along a path is not universal, as an Ornstein--Uhlenbeck process with correlation time $\tau_c=\beta\,\tau_K$ and take $\beta=1$, a single time although vorticity decorrelates more slowly than strain \citep{Yeung1989}. The model concerns $\mathbf f$ at a fixed orientation. When $\E$ and $\W$ are themselves evolved as Ornstein--Uhlenbeck tensors sharing that time, as in the tensor-forced simulations (Appendix~\ref{app:numerics}), $\mathbf f$ at fixed $\p$ is a linear combination of their components and so exactly Ornstein--Uhlenbeck; along a path Eq.~\eqref{eq:fdef} is evaluated at the moving orientation and $\mathbf f$ is no longer strictly so, a difference the closure neglects and the simulations retain (Secs.~\ref{sec:watson} and~\ref{sec:conditioning}). Settling faster than the eddies decorrelate shortens the correlation to the crossing time $\ell/w$ \citep{Gustavsson2019,Gustavsson2021}, which falls below $\tau_K$ once $\Sv$ exceeds $\ell/\eta\simeq12$--$20$, according to shape and Reynolds number (Appendix~\ref{app:tauc}), well above the slip maximum studied here. The value of $\beta$ cancels from the aligned limit (Sec.~\ref{sec:correspondence}) and moves the peak constant $c^*$ of Sec.~\ref{sec:peak} for the $\kappa=10$ column from $0.925$ at $\beta=1$ to $0.884$ at $\beta=1/2$ and $0.956$ at $\beta=2$.

\section{Stationary orientation statistics}
\label{sec:stationary}

\subsection{Coloured-noise linear response}
\label{sec:linresp}

Write $\bm\delta$ for the small tangential displacement of $\p$ from the attracting set. Linearising the drift, each restored component obeys $\dot\delta=-\lambda_r\delta+f(t)$, with Ornstein--Uhlenbeck forcing of per-component variance $\sigma_f^2$ from Eq.~\eqref{eq:fvar} and correlation time $\tau_c$. Since $|\p|=1$, the tilt equals the displacement at leading order: $\varphi^2=\delta^2$ for a girdle, with one restored direction, and $\varphi^2=\delta_1^2+\delta_2^2$ for a bipolar state, with two. The stationary variance of this linear problem is exact,
\begin{equation}
\langle\delta^2\rangle
=\frac{\sigma_f^2\,\tau_c}{\lambda_r\,(1+\lambda_r\tau_c)} ,
\label{eq:linresp}
\end{equation}
which with $\tau_c=\tau_K$ and Eq.~\eqref{eq:lamr} gives, per restored component,
\begin{equation}
\langle\delta^2\rangle
=\frac{5+3\mathcal{B}^2}{60\,F\Sv^2\,(1+F\Sv^2)} .
\label{eq:gauss}
\end{equation}
Equation~\eqref{eq:gauss} recovers the strong-alignment scalings of the rapid-settling theories \citep{Menon2017,Roy2023,Gustavsson2019} and identifies the coloured-noise correction $(1+F\Sv^2)^{-1}$: the effective rotational diffusivity is
\begin{equation}
D_{\rm eff}=\frac{\sigma_f^2\,\tau_c}{1+\lambda_r\tau_c},
\label{eq:Deff}
\end{equation}
reduced below the white-noise value $D=\sigma_f^2\tau_c$ by the factor $1+\lambda_r\tau_c$, substantially once the relaxation is faster than the decorrelation of the forcing.

\subsection{The Boltzmann density}
\label{sec:watson}

For Gaussian white-noise forcing with isotropic statistics the full nonlinear problem is isotropic rotational diffusion on the sphere, with diffusivity $D$, in the gradient drift of Sec.~\ref{sec:gradflow}. That drift gives $\dot\mu=-s_\kappa\lambda_r\mu(1-\mu^2)$, so a density $\psi(\mu,t)$ uniform in azimuth obeys
\begin{equation*}
\partial_t\psi=\partial_\mu\bigl\{(1-\mu^2)\bigl[s_\kappa\lambda_r\mu\,\psi+D\,\partial_\mu\psi\bigr]\bigr\},
\end{equation*}
and its stationary solution, which carries no probability current, is of Boltzmann form, $\psi\propto\exp[-V(\mu)/D]$, exactly and at every $\Sv$; no small-tilt assumption enters, only the gradient structure. With the quadratic potential it is the Watson distribution of directional statistics \citep{Watson1965,Mardia2000}, which we call the Boltzmann density,
\begin{equation}
\psi(\mu)=\frac{e^{\mp b\,\mu^2}}{Z_\mp},
\qquad
\begin{cases} -\;:\;\text{prolate (girdle)},\\ +\;:\;\text{oblate (bipolar)},\end{cases}
\label{eq:watson}
\end{equation}
with concentration $b=\lambda_r/2D$. Here and below $\psi$ is the density of $m\equiv|\mu|=|\p\cdot\ghat|$ on $[0,1]$, the density per unit solid angle being $\psi/4\pi$. Since $\psi$ depends on $\mu^2$ alone we write it in $\mu$, and statistics involving the folded variable, such as $\langle(1-m)^2\rangle$ or the tilt $\arccos m$ of a plate, are written in $m$. The normalisations are
\begin{equation}
Z_-=\int_0^1\!e^{-b\mu^2}\dd\mu=\frac12\sqrt{\frac\pi b}\,\mathrm{erf}\sqrt b,
\qquad
Z_+=\int_0^1\!e^{b\mu^2}\dd\mu=\frac{e^{b}\mathcal D(\sqrt b)}{\sqrt b},
\label{eq:Znorm}
\end{equation}
with $\mathcal D(x)=e^{-x^2}\!\int_0^x e^{t^2}\dd t$ the Dawson function. The orientation performs an overdamped random walk in a quadratic well, with $\lambda_r$ and $D$ in the roles of the spring constant and the temperature, each divided by the friction; confinement to the sphere turns the Gaussian of a harmonic well into the density \eqref{eq:watson} and makes it saturate at the uniform measure as $b\to0$. Unlike thermal noise, the turbulent forcing obeys no fluctuation--dissipation relation: $D$ is set by the velocity gradients and, within the prescribed forcing model, is independent of the particle's size and rotational friction, while $\lambda_r$ grows with the settling speed, so the particle size alone changes $b$ by orders of magnitude.

For forcing of finite correlation time we replace $D$ by $D_{\rm eff}$, so that
\begin{equation}
b \;=\; \frac{\lambda_r}{2D_{\rm eff}}
\;=\; \frac{30\,F\Sv^2\,(1+F\Sv^2)}{5+3\mathcal{B}^2},
\label{eq:b}
\end{equation}
or $b=30F\Sv^2(1+\beta F\Sv^2)/[\beta(5+3\mathcal{B}^2)]$ for general $\beta$. This replacement is a finite-memory closure, tested below, and the only approximation beyond the forcing assumptions of Sec.~\ref{sec:forcing}: the Boltzmann form, exact for isotropic Gaussian white noise, is kept and its one parameter is matched to the exact linear response. A systematic coloured-noise elimination of the Jung--H\"anggi type \citep{Jung1987,Hanggi1995}, exact for a linear drift, would not give a density of this form here, since the restoring drift $\lambda_r\mu(1-\mu^2)$ is cubic in the tangent coordinate; it would produce an orientation-dependent prefactor and higher harmonics in the exponent. Equation~\eqref{eq:b} nevertheless has the correct limits. As settling vanishes at fixed shape, statistical isotropy makes the density uniform whatever the forcing (Sec.~\ref{sec:forcing}), and Eq.~\eqref{eq:b} returns it; switching off the torque while the particle keeps settling is a different limit, and the inertial runs of \cite{Anand2020} with the torque suppressed are not uniform (insets of their Fig.~1). A dominant torque keeps the tilt small, where the linear response \eqref{eq:linresp} that supplies $D_{\rm eff}$ is exact for forcing with the fixed-orientation statistics, so the closure recovers the aligned second-moment coefficient (Sec.~\ref{sec:correspondence}); that assumption is controlled there for plates but not for columns (Sec.~\ref{sec:forcing}).

In between, we measure the accuracy of the closure by integrating the equation it describes,
\begin{equation}
\dot{\p}=\pm\lambda_r\mu(\I-\p\p)\cdot\ghat+\mathbf f(t),
\label{eq:langevin}
\end{equation}
with the upper sign for plates and the lower for columns, for two design shapes, a prolate spheroid of aspect ratio $\kappa=10$ and an oblate one of $\kappa=0.03$, at the reference value $F=0.316$ with $\Sv=\Sv_{\rm eq}$. Their physical coefficients are $F_{\rm eq}=0.060$ and $0.043$, but orientation moments depend on $F\Sv^2$ alone, so the curves apply to the physical particles after the rescaling $\Sv_{\rm eq}\to\Sv_{\rm eq}(0.316/F_{\rm eq})^{1/2}$, which the root-mean-square (rms) slip, carrying an explicit $\Sv_{\rm eq}$, needs in its ordinate as well. An ambient vector $\bm\xi(t)$, independent of the particle, is advanced as a three-component Ornstein--Uhlenbeck process with correlation time $\tau_c$ and per-component variance $\sigma_f^2$, and $\mathbf f=(\I-\p\p)\cdot\bm\xi$ is taken at the current orientation. At a fixed orientation this reproduces Eq.~\eqref{eq:fvar}, though not along a trajectory (Sec.~\ref{sec:conditioning}). Unlike the Jeffery forcing \eqref{eq:fdef}, which is odd under $\p\to-\p$ as a director requires, $(\I-\p\p)\cdot\bm\xi$ is even in each realisation; its ensemble statistics are even, and the tensor-forced runs below give the same moments to within $1$--$2.5\%$. The velocity gradient itself is not simulated, so the test probes the closure within the forcing model rather than the model itself.

\begin{figure}
\includegraphics[width=\textwidth]{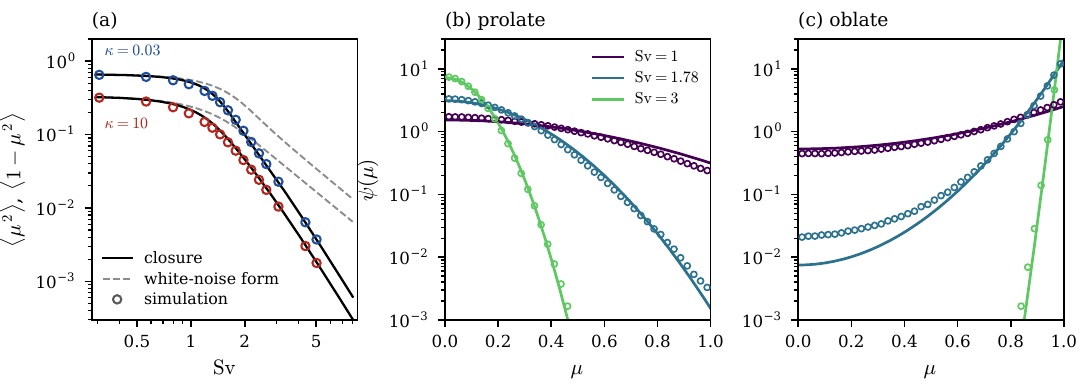}
\caption{Tests of the closure for the column ($\kappa=10$) and the plate ($\kappa=0.03$) at $F=0.316$. (a) $\langle\mu^2\rangle$ for the column and $\langle1-\mu^2\rangle$ for the plate: solid, the closure \eqref{eq:b}; dashed, the same without the factor $1+F\Sv^2$; symbols, simulations with projected-vector forcing. (b), (c) Stationary densities at three settling numbers from tensor-forced simulations (symbols), against Eq.~\eqref{eq:watson} at the concentration \eqref{eq:b}.}
\label{fig:sde}
\end{figure}

Figure~\ref{fig:sde}(a) shows $\langle\mu^2\rangle$ for the column and $\langle1-\mu^2\rangle$ for the plate. The closure tracks the simulations across more than two decades, with its largest errors in the crossover (plotted in Sec.~\ref{sec:peak}). It overpredicts the prolate $\langle\mu^2\rangle$ by up to $15\%$, near $F\Sv^2\approx0.45$. The oblate error changes sign: the closure overpredicts the spread by up to $8\%$ near $F\Sv^2\approx0.4$ and underpredicts it by up to $6.5\%$ near $F\Sv^2\approx1$, and the error falls to $1\%$ by $F\Sv^2=3$. The prolate error decays more slowly, to $5\%$ at $F\Sv^2=3$ and $3\%$ at $6$. Under the projected-vector forcing a column's azimuth is free, and the particle turns towards the horizontal part of the ambient vector, which then pulls it towards the horizontal as well and adds to the restoring rate at any settling number; freezing the azimuth in these simulations removes the deficit, leaving an error of $1$--$3\%$ of the opposite sign (Appendix~\ref{app:numerics}). That mechanism belongs to the vector model, whereas the weakening of the forcing felt along a path appears under both forcings (Sec.~\ref{sec:conditioning}). At the smallest settling numbers tested, $F\Sv^2=0.03$ and $0.1$, the error in these second moments is only $1$--$6\%$, but the departure from the uniform density, $|\langle\mu^2\rangle-\tfrac13|$, is about $1.5$ times that of the closure for the column and $1.5$--$1.6$ times for the plate, and from $F\Sv^2=0.1$ to $0.03$ the ratio shows no resolved decrease towards unity. We attribute this to the ballistic motion below $\tau_c$: the forcing turns the orientation through an angle of order $\sigma_f\tau_c\approx0.36$ before it decorrelates, and a weak drift then shifts the density more than diffusion of the same strength would. The closure thus has the uniform limit right but, at the torques tested, not the approach to it, and on this argument the discrepancy should grow with $\sigma_f\tau_c$; the plate simulations of Sec.~\ref{sec:anand} show a factor of $1.6$--$1.9$ at their weakest torques. Without the coloured-noise factor $1+F\Sv^2$ the concentration would be too low by a factor of two at $\Sv=1.78$ and nine at $\Sv=5$ (dashed curves).

Panels (b) and (c) compare the full stationary densities. The forcing there is built from Ornstein--Uhlenbeck tensors $\E$ and $\W$ with the covariances \eqref{eq:Ecov}--\eqref{eq:Wcov}, and $\mathbf f$ is evaluated from Eq.~\eqref{eq:fdef}, so neither the projection nor the tangential isotropy is imposed by hand; the two constructions agree on the variance to within $1\%$ for the column and $2.5\%$ for the plate, so the closure errors above carry over, the prolate maximum becoming $16\%$ (Appendix~\ref{app:numerics}). The densities follow Eq.~\eqref{eq:watson} across the crossover, the largest departure being in the far tail of the oblate case at $\Sv=1.78$, where the measured density lies above the closure. The forcing in these tests is Gaussian; resolved turbulence is intermittent, which reshapes the density (Sec.~\ref{sec:intermittency}). Section~\ref{sec:anand} tests the result against direct numerical simulation.

\subsection{Limiting forms and the rapid-settling coefficient}
\label{sec:moments}
\label{sec:correspondence}

\begin{figure}
\includegraphics[width=\textwidth]{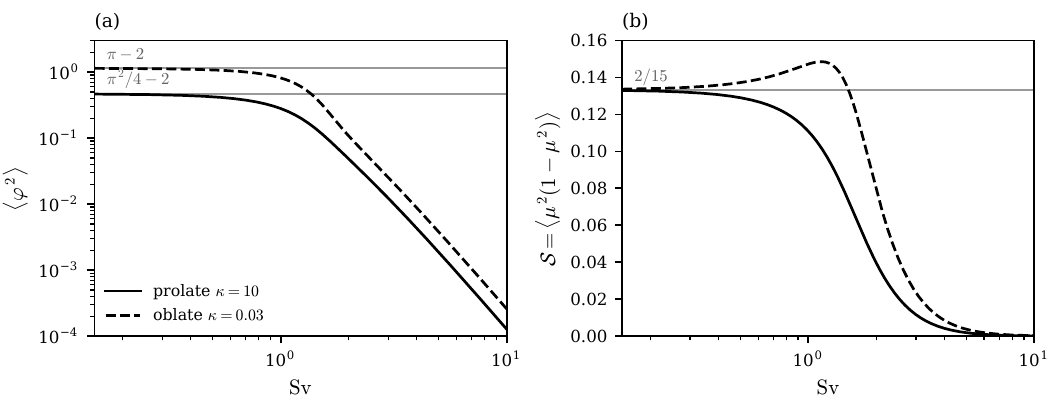}
\caption{Moments of the Boltzmann density at $F=0.316$ (solid, the column $\kappa=10$; dashed, the plate $\kappa=0.03$). (a) Mean-square tilt, with the random-orientation limits \eqref{eq:randpro}. (b) The slip shape factor $\mathcal S$ of Eq.~\eqref{eq:vh2}.}
\label{fig:moments}
\end{figure}

Every even moment of $\mu$ under Eq.~\eqref{eq:watson} is available in closed form (Appendix~\ref{app:moments}), whereas angle moments such as $\langle\varphi^2\rangle$ need a quadrature (figure~\ref{fig:moments}a). As $b\to0$ the mean-square tilt saturates at the uniform-measure values
\begin{equation}
\langle\varphi^2\rangle_{\rm pro}\to\frac{\pi^2}{4}-2\simeq0.4674,
\qquad
\langle\varphi^2\rangle_{\rm obl}\to\pi-2\simeq1.1416,
\label{eq:randpro}
\end{equation}
and once aligned it falls as
\begin{equation}
\langle\varphi^2\rangle\simeq\frac{5+3\mathcal{B}^2}{60\,F^2\Sv^4}
\quad\text{(girdle)},
\qquad
\text{twice that for a bipolar state},
\label{eq:asymp}
\end{equation}
the $\Sv^{-4}$ law of the rapid-settling theories \citep{Menon2017,Roy2023,Gustavsson2019}. Equation~\eqref{eq:asymp} drops the $1$ in $1+F\Sv^2$ as well as taking the small-tilt limit, and the first causes most of the error: for the $\kappa=10$ column at $F=0.316$ its error is $85\%$ at $\Sv=1.78$ and still $9\%$ at $\Sv=6$, whereas the small-tilt result $1/2b$ with the full concentration \eqref{eq:b} is only $7\%$ low at $\Sv=1.78$.

In the aligned limit $\lambda_r\tau_c\gg1$, Eq.~\eqref{eq:b} gives $b\to\lambda_r^2/2\sigma_f^2$, with $\tau_c$ cancelling between the diffusivity and the coloured-noise factor, so the limit carries no modelling parameter and can be compared directly with the rapid-settling theories. Following \cite{Gustavsson2019,Gustavsson2021}, write the relaxation rate as $A_\tau\Sv_{\rm G}^2$, with $A_\tau(\kappa)$ their dimensionless torque coefficient, $\Sv_{\rm G}$ a settling number built on the particle response time and $A_g(\kappa)$ the translational resistance along gravity, so that $w/u_\eta=\Sv_{\rm G}/A_g$ and $F=A_\tau A_g^2$. Equation~\eqref{eq:asymp} then becomes $\langle\varphi^2\rangle=C(\kappa)\,\varepsilon\nu/w^4$, with
\begin{equation}
C(\kappa)=\frac{d_\perp\,(5+3\mathcal{B}^2)}{60\,A_\tau^2A_g^4},
\qquad
d_\perp=\begin{cases}1, & \kappa>1,\\ 2, & \kappa<1,\end{cases}
\label{eq:Ckappa}
\end{equation}
where $d_\perp$ counts the restored directions (Sec.~\ref{sec:topology}). This is the shape coefficient of the rapid-settling theory, obtained in the slender-body limit by \cite{Menon2017} and \cite{Roy2023} and for arbitrary aspect ratio by \cite{Gustavsson2019,Gustavsson2021}, recovered here under the same fixed-orientation forcing. A white-noise branch arises only if crossing sets $\tau_c=\ell/w$ with $\lambda_r\tau_c\ll1$. Equation~\eqref{eq:linresp} then gives $\langle\varphi^2\rangle\propto\Sv^{-3}$, the exponent of the many-eddy limit of a random field traversed at large mean speed \citep{Kesten1979}, applied to settling pairs by \cite{patra2026collisionefficiencyrapidlysettling}, and the one with which \cite{Gustavsson2021} corrected the $\Sv^{-2}$ of \cite{Klett1995}; it needs $\ell/\eta\ll\Sv\ll\eta/(F\ell)$, hence $F(\ell/\eta)^2\ll1$, which only nearly spherical particles satisfy.

The density of \cite{Roy2023} describes the persistent end of the same dynamics. They push a shear-rate distribution, from data or from a stochastic gradient model \citep{Girimaji1990}, through the equilibrium \eqref{eq:maptan}, the limit $\lambda_r\tau_c\gg1$ in which the tilt follows the forcing instantaneously; the Boltzmann form is exact at the white-noise end, and Eq.~\eqref{eq:b} carries its concentration between the two. The densities coincide near alignment, where Eq.~\eqref{eq:maptan} linearises to $\delta=f/\lambda_r$, and part at large tilt, the map giving $\sec^2\varphi\,\exp(-\lambda_r^2\tan^2\varphi/2\sigma_\gamma^2)$ for Gaussian shear of variance $\sigma_\gamma^2$ against the Boltzmann form $\exp(-b\sin^2\varphi)$. The map also requires an equilibrium at each value of the forcing. A disc in vertical shear always has one, and so does a particle under the frozen projected-vector forcing of Sec.~\ref{sec:watson}, which adds the gradient of $\bm\xi\cdot\p$ to the drift and leaves a gradient flow on the sphere. A column turned by a rigid rotation of rate $\Omega$ about the horizontal axis normal to it obeys $\dot\varphi=\Omega-\tfrac12\lambda_r\sin2\varphi$ and loses its equilibrium once $|\Omega|>\tfrac12\lambda_r$, as under strong vortical forcing; the Fokker--Planck description does not require one.

\subsection{A worked example}

Take the two design shapes at $F\Sv^2\simeq1$, in the middle of the crossover: $\Sv=1.78$ at the reference value $F=0.316$, or $\Sv_{\rm eq}\simeq4.07$ for the physical column and $4.83$ for the physical plate (Sec.~\ref{sec:watson}). Equation~\eqref{eq:b} gives $b=7.63$ for the column and $7.52$ for the plate, and the moments follow at once: $\langle\mu^2\rangle=0.065$ and $0.852$. A Rayleigh scattering calculation needs the orientation average of the squared projection of $\p$ on a polarisation direction, which for a horizontal direction is $\tfrac12(1-\langle\mu^2\rangle)$: $0.467$ for the column and $0.074$ for the plate. Random orientation gives $1/3$, and perfect broadside alignment $1/2$ or $0$. Neither idealisation is close for the plate, and the column lies between the two, nearer the aligned value. A calculation weighted towards large tilts also depends on the tails, which in turbulence differ from those of a single Boltzmann density (Sec.~\ref{sec:intermittency}).

\section{Slip-velocity statistics}
\label{sec:slip}

With the quasi-static slip \eqref{eq:slipvec} in Kolmogorov units, $\w/u_\eta=\Sv(\hat U_\perp\ghat+g_M\mu\p)$ with $\Sv=\Sv_{\rm eq}$, the vertical and lateral components are
\begin{equation}
v_z=\Sv\bigl(\hat U_\perp+g_M\mu^2\bigr),
\qquad
\mathbf v_h=\Sv\,g_M\,\mu\sqrt{1-\mu^2}\;\hat{\mathbf e}_h ,
\label{eq:vzvh}
\end{equation}
with $\hat{\mathbf e}_h$ the unit vector along the horizontal projection of $\p$. For an axisymmetric density, the azimuth being uniform,
\begin{equation}
\langle v_z\rangle = \Sv\bigl(\hat U_\perp+g_M\langle\mu^2\rangle\bigr),
\qquad
\langle|\mathbf v_h|^2\rangle = \Sv^2 g_M^2\,\mathcal S ,
\qquad
\mathcal S\equiv\langle\mu^2(1-\mu^2)\rangle ,
\label{eq:vh2}
\end{equation}
and $\langle\mathbf v_h\rangle=\mathbf 0$. Here $\langle v_z\rangle$ is the mean vertical slip; the mean settling velocity adds the fluid velocity sampled along the path, biased downwards by preferential sweeping and absent from the overdamped slip. Equations~\eqref{eq:vzvh} and \eqref{eq:vh2} show that all orientation dependence enters through the mobility contrast $g_M$, which vanishes for a sphere, and that the lateral slip is set by a fourth orientation moment, which makes it more sensitive to the tails of the density than the settling speed.

The lateral slip is odd in the tilt, so its rms is first order in the tilt spread, whereas the vertical fluctuation is even and second order: near alignment a particle wanders sideways before its fall speed changes measurably. In the random limit $\mathcal S\to2/15$ and $\mathrm{var}(v_z)/\langle|\mathbf v_h|^2\rangle\to2/3$; once aligned $\mathcal S$ collapses onto the mean-square tilt, so that $\langle|\mathbf v_h|^2\rangle\propto\Sv^{-2}$ while $\mathrm{var}(v_z)\propto\Sv^{-6}$. The same coupling through $g_M$, driven by a deterministic wave field rather than by turbulence, sets the drift of spheroids settling through surface gravity waves \citep{MishraWaves2025}.

For the Boltzmann density $\mathcal S$ is elementary (Appendix~\ref{app:moments}). It runs from $2/15$ at random orientation to $1/2b$ for a girdle and $1/b$ for a bipolar state once aligned, so that
\begin{equation}
\langle|\mathbf v_h|^2\rangle^{1/2}=\Sv\,|g_M|\,\mathcal S^{1/2}
\label{eq:vhrms}
\end{equation}
is available at any settling number without quadrature. Figure~\ref{fig:moments}(b) shows $\mathcal S$; for plates it overshoots $2/15$ at intermediate $\Sv$, because, as $b$ grows, weight moves from the equator towards the poles through $\mu^2\approx\tfrac12$, where $\mu^2(1-\mu^2)$ is largest.

\section{Maximum of the instantaneous rms lateral slip}
\label{sec:peak}

The lateral slip is a fluctuation, $\langle\mathbf v_h\rangle=\mathbf 0$, and Eq.~\eqref{eq:vhrms} is its root-mean-square. It vanishes at both ends, through the amplitude $\Sv$ at small $\Sv$ and through the loss of tilt at large $\Sv$, so it peaks in between, at $\Sv^*$. The maximum is over the settling speed in a fixed flow, for instance over particle size at fixed aspect ratio, and not over the dissipation rate for a fixed particle, whose dimensional slip $W_{\rm eq}|g_M|\mathcal S^{1/2}$ simply follows $\mathcal S(b)$ (figure~\ref{fig:moments}b). Writing $u=F\Sv^2$, so that $b=u(1+u)/2\sigma_f^2\tau_K^2$, stationarity of $\Sv^2\mathcal S$ requires
\begin{equation}
\frac{\dd\ln\mathcal S}{\dd\ln b}=-\,\frac{1+u}{1+2u} .
\label{eq:peakcond}
\end{equation}
The right-hand side lies between $-1$ and $-\tfrac12$ for any shape, and $\mathcal S$ depends on $b$ alone, so $b$ at the peak falls in a narrow window: $b^*=6.03$--$6.35$ for prolate shapes from $\kappa=2$ to $20$ and $7.47$--$7.75$ for oblate ones from $\kappa=0.5$ to $0.03$. Shape then enters only through the forcing variance,
\begin{equation}
u^*(1+u^*)=\frac{(5+3\mathcal{B}^2)\,b^*}{30}.
\label{eq:ustar}
\end{equation}
The peak therefore depends only weakly on shape: $5+3\mathcal{B}^2$ runs from $5$ to $8$ between a sphere and a slender rod or thin disc, and $c^*=\sqrt{u^*}$ varies roughly as its cube root, so that
\begin{equation}
\Sv^*\;=\;\frac{c^*}{\sqrt F},
\qquad c^*\in[0.86,\,1.00],
\label{eq:svstar}
\end{equation}
from $0.860$ at $\kappa=2$ to $0.998$ at $\kappa=0.03$, while the torque coefficient $F_b$ itself rises from $0.12$ to $0.32$ (Appendix~\ref{app:mobility}). The forcing statistics move $c^*$ by as much as the shape does (see below). Nearer the sphere $c^*$ tends to $0.814$ from the prolate side and $0.870$ from the oblate side, since the attracting set remains a girdle or a pole pair however small $|\mathcal{B}|$ becomes, but there $g_M$ and $F$ vanish and the peak is unobservable. Elsewhere, locating the peak gives the torque coefficient,
\begin{equation}
F=\bigl(c^*/\Sv^*\bigr)^2 .
\label{eq:Finv}
\end{equation}

\begin{figure}
\includegraphics[width=\textwidth]{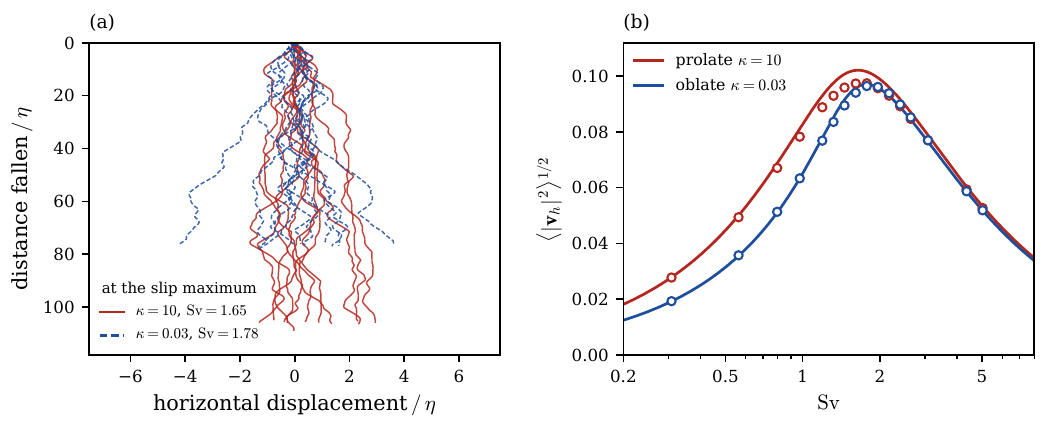}
\caption{Lateral slip of the two design shapes at $F=0.316$. (a) Slip-only paths in the vertical plane at the peak, eleven per shape from one release point. (b) Root-mean-square lateral slip in Kolmogorov units: curves, Eq.~\eqref{eq:vhrms}; circles, the simulations of figure~\ref{fig:sde}(a).}
\label{fig:drift}
\end{figure}

Figure~\ref{fig:drift} illustrates the maximum. Integrating the orientation equation together with the particle position, Eq.~\eqref{eq:vzvh}, without the displacement by the carrier fluid, gives paths that wander sideways by a few Kolmogorov lengths over a hundred lengths of fall (panel a), and the rms slip in the simulations of figure~\ref{fig:sde}(a) follows Eq.~\eqref{eq:vhrms} and peaks close to the settling number of Eq.~\eqref{eq:svstar} (panel b).

\begin{figure}
\includegraphics[width=\textwidth]{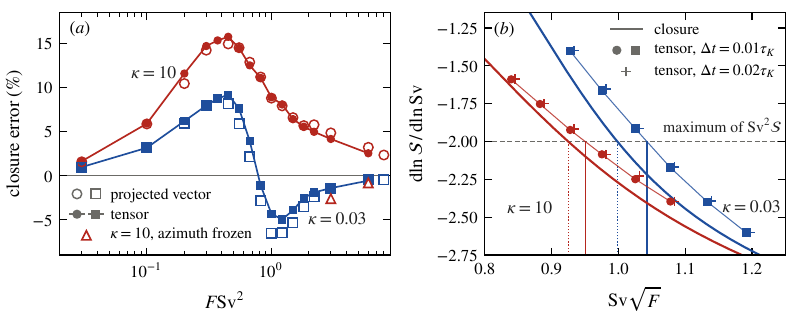}
\caption{Closure accuracy and the slip maximum for the $\kappa=10$ column (circles) and the $\kappa=0.03$ plate (squares). (a) Closure error in the spread, $100\,(X_{\rm cl}/X_{\rm sim}-1)$, with $X=\langle\mu^2\rangle$ for the column and $\langle1-\mu^2\rangle$ for the plate: open symbols, projected-vector forcing; filled, tensor forcing; triangles, the column with its azimuth frozen. (b) The slope $\dd\ln\mathcal S/\dd\ln\Sv$, whose value $-2$ (dashed) marks the maximum, at $\Sv\sqrt F=c^*$: lines, the closure; symbols, tensor-forced simulations at $\Delta t=0.01\tau_K$ (filled) and $0.02\tau_K$ (crosses); vertical lines, $c^*$ from the closure (dotted) and the simulations (solid).}
\label{fig:peaktest}
\end{figure}

Figure~\ref{fig:peaktest}(a) collects the second-moment test of Sec.~\ref{sec:watson}: the closure error in the spread about the attracting set under both forcings, and for the column with its azimuth frozen. The inversion below rests on the location of the maximum, which that test does not establish, since $c^*$ is set by the logarithmic slope of $\mathcal S$, a fourth-moment quantity, through the peak. We therefore located the maximum of $u\mathcal S$ in tensor-forced simulations of the two design shapes, on seven values of $u$ spaced by $0.1$ in $\ln u$ and driven by common random numbers to resolve that slope (Appendix~\ref{app:numerics}). They give $c^*=0.950\pm0.002$ for the column and $1.042\pm0.002$ for the plate, against $0.925$ and $0.998$ from the closure (figure~\ref{fig:peaktest}b); the uncertainty covers sampling and the choice of fit, and doubling the time step moves either value by at most $0.002$. Across the grid the closure overestimates $\mathcal S$ by $10\%$ falling to $7\%$ for the column, and by $4\%$ for the plate, turning into a $1\%$ underestimate: the error decreases through the peak, towards the aligned limit, so the simulated $\mathcal S$ decays more slowly and $\Sv^2\mathcal S$ peaks $3\%$ and $4\%$ higher in $\Sv$. Taken at its closure value, $c^*$ therefore underestimates $F$ in Eq.~\eqref{eq:Finv} by $5\%$ and $8\%$ and overestimates $\varepsilon$ in Eq.~\eqref{eq:epsinv} by $11\%$ and $19\%$. Since $b$ at the peak varies little within each family, similar corrections should apply to the other aspect ratios of Eq.~\eqref{eq:svstar}, though the simulations cover only these two shapes.

The location of the peak also depends on the forcing actually felt. If that is a fraction $r$ of the fixed-orientation variance \eqref{eq:fvar}, lowering $r$ from $1$ to $0.6$ lowers $c^*$ from $0.925$ to $0.80$ for the column and from $0.998$ to $0.86$ for the plate, so the fixed-orientation value overstates $F$ by $33\%$ and $35\%$. The peak therefore constrains $F$ and the forcing statistics together: with $r$ calibrated it returns $F$, and with $F$ computed from the shape it returns $r$, a scalar measure of the sampling bias of Sec.~\ref{sec:conditioning}. Dissipation intermittency raises $c^*$ by up to $11\%$ for the column and $14\%$ for the plate (Sec.~\ref{sec:intermittency}).

\subsection{Inverting the peak}
\label{sec:invert}

If the shape and size are known, so is $F$, and the peak constrains the flow instead. From $\Sv^*=c^*/\sqrt F$ the Kolmogorov velocity is $u_\eta=w^*\sqrt F/c^*$, with $w^*$ the settling speed at the peak in the convention in which $F$ is quoted, so that
\begin{equation}
\varepsilon=\frac{F^2\,w^{*4}}{\nu\,c^{*4}},\qquad
\eta=\frac{\nu}{u_\eta},\qquad
\tau_K=\frac{\nu}{u_\eta^2}.
\label{eq:epsinv}
\end{equation}
The inversion amplifies the uncertainty in $F$ and in the forcing statistics: since $\varepsilon\propto F^2c^{*-4}$, a $7\%$ error in $c^*$ becomes $31\%$ in $\varepsilon$. For $\kappa=10$ columns in cirrus with $\varepsilon=10^{-4}\,\mathrm{m^2\,s^{-3}}$ and $\nu=1.5\times10^{-5}\,\mathrm{m^2\,s^{-1}}$, the peak falls at a broadside speed $w^*\simeq1.3\,\mathrm{cm\,s^{-1}}$ ($F_b=0.189$, $\Sv_b^*=2.13$), or an equivalent-sphere speed of $2.3\,\mathrm{cm\,s^{-1}}$ ($F_{\rm eq}=0.060$), reached by ice columns about $0.14\,$mm long; at $\varepsilon=10^{-2}\,\mathrm{m^2\,s^{-3}}$ it moves to $4.2\,\mathrm{cm\,s^{-1}}$ and columns about $0.24\,$mm long. Whether such crystals are overdamped is examined in Sec.~\ref{sec:application}; in air the condition is restrictive.

Both readings use the slip speed, not a displacement, and the lateral dispersion need not peak where the slip does, since it depends also on the correlation time of $\mathbf v_h$. For an aligned plate the lateral slip follows the two restored tilt components, whose integral time is $\tau_c+\lambda_r^{-1}$ by the linear response \eqref{eq:linresp}; for a column $\mathbf v_h\propto\mu\,\hat{\mathbf e}_h$, and the horizontal direction $\hat{\mathbf e}_h$ keeps diffusing in azimuth at any settling number, so its decorrelation enters as well. A displacement also carries the carrier-fluid velocity and its correlation with the slip, which differencing against a sphere does not remove, the two particles following different paths. We therefore state the result as a maximum of the instantaneous slip.

Because the maximum is flat, its position is hard to measure precisely, but the closed form can be used at any settling number. The ratio of the lateral rms to the mean vertical slip,
\begin{equation}
\mathcal R_v\equiv\frac{\langle|\mathbf v_h|^2\rangle^{1/2}}{\langle v_z\rangle}
=\frac{|g_M|\,\mathcal S^{1/2}}{\hat U_\perp+g_M\langle\mu^2\rangle},
\label{eq:ratio}
\end{equation}
is dimensionless, needs no knowledge of $u_\eta$, and depends only on $\Sv$ once the shape and the forcing model are fixed. For columns with $\hat U_\parallel/\hat U_\perp<38/23$, that is $\kappa\lesssim95$ (Appendix~\ref{app:moments}), which includes the range $2\le\kappa\le20$ treated here, it falls monotonically, from $|g_M|(2/15)^{1/2}/(\hat U_\perp+g_M/3)$ at random orientation to a slope near $-2$ once aligned, so one measurement of $\mathcal R_v$ fixes $\Sv$, and $u_\eta$ follows from the settling speed; for more slender columns $\mathcal R_v$ rises by less than $1\%$ before falling. For plates $\mathcal R_v$ first rises, by about $18\%$ to a maximum near $F\Sv^2=0.62$ at $\kappa=0.03$ ($\Sv_b\simeq1.4$, $\Sv_{\rm eq}\simeq3.8$), and only then falls, so a second observable must select the branch.

With $F$ computed from the shape (Appendix~\ref{app:mobility}), Eq.~\eqref{eq:svstar} places the peak at $\Sv_{\rm eq}^*\simeq3.8$ for the $\kappa=10$ column and $4.8$ for the plate, or $\Sv_b^*\simeq2.1$ and $1.8$ on the broadside speed ($F_b\simeq0.19$ and $0.32$), and at $\Sv_b^*\simeq3.0$ and $2.5$ with the finite-$\Rep$ factor $C_\tau=0.5$ of Appendix~\ref{app:mobility}.

\section{Extensions}
\label{sec:beyond}

\subsection{General torques}
\label{sec:aniso}

Since any axisymmetric drift $g(\mu)(\I-\p\p)\cdot\ghat$ is a gradient flow (Sec.~\ref{sec:gradflow}), white forcing with isotropic statistics gives $\psi\propto e^{-V(\mu)/D}$ exactly for any such torque. Higher harmonics from finite-$\Rep$ corrections, nonlinear drag entering through an orientation-dependent slip, and superposed body-force torques are all of this form, and the density of Sec.~\ref{sec:watson} is the quadratic member of the class. For coloured forcing the closure of Sec.~\ref{sec:watson} carries over only locally, through the relaxation rate about an isolated attractor.

The Boltzmann form also needs forcing that is isotropic in the tangent plane. Forcing that distinguishes a direction, as stratified or sheared turbulence may, makes the noise depend on orientation, and the stationary density then departs from that form. Settling alone produces such anisotropy, through the crossing integral of Appendix~\ref{app:tauc}, only in the white-noise branch, which only nearly spherical particles reach (Sec.~\ref{sec:correspondence}); in the persistent regime the tilt follows the instantaneous forcing, which is isotropic in the tangent plane.

\subsection{Intermittent dissipation}
\label{sec:intermittency}

The forcing so far is Gaussian, with a variance set by the mean dissipation, but turbulent velocity gradients are intermittent: the local pseudo-dissipation fluctuates over orders of magnitude, its logarithm is close to normal \citep{Yeung1989,Pope2000}, and its variance grows with the Taylor-scale Reynolds number \citep{Kolmogorov1962,Sreenivasan1997}. We use the parametrisation of \cite{Koch2002},
\begin{equation}
\sigma^2_{\ln\bar\phi}=-0.15+0.25\ln\bigl(3.1\,\mathrm{Re}_\lambda^{1/2}+0.1\,\mathrm{Re}_\lambda^{3/2}\bigr),
\label{eq:sigmaRe}
\end{equation}
and represent intermittency by a scalar amplitude: $\varepsilon_{\rm loc}=\bar\phi\,\langle\varepsilon\rangle$, with $\ln\bar\phi$ normal of variance $\sigma^2_{\ln\bar\phi}$ and mean $-\tfrac12\sigma^2_{\ln\bar\phi}$, so that $\langle\bar\phi\rangle=1$. This models the amplitude of the forcing, not the full gradient tensor conditioned on the true dissipation. The restoring rate $\lambda_r$ is fixed by the settling speed and does not fluctuate, whereas the forcing variance does, $\sigma_f^2\to\bar\phi\,\sigma_f^2$. In the aligned limit, where $b\to\lambda_r^2/2\sigma_f^2$ and the correlation time has dropped out, the local concentration is therefore
\begin{equation}
b(\bar\phi)=b_0/\bar\phi,\qquad b_0=\lambda_r^2/2\sigma_f^2 .
\label{eq:blocal}
\end{equation}
The same form holds at any settling number, with $b_0$ replaced by the closure concentration \eqref{eq:b}, if the correlation time does not follow the local dissipation. If the orientation equilibrates faster than $\bar\phi$ evolves, the stationary density is the lognormal average of Boltzmann densities over an effective diffusivity that follows the local dissipation, a superstatistics in the sense of \cite{Beck2003},
\begin{equation}
\psi(\mu)\;=\;\int_0^\infty\!\dd\bar\phi\;P(\bar\phi)\,
\frac{e^{\mp b_0\mu^2/\bar\phi}}{Z_\mp(b_0/\bar\phi)},
\qquad
P(\bar\phi)=\frac{1}{\bar\phi\sqrt{2\pi\sigma^2_{\ln\bar\phi}}}
\exp\!\left[-\frac{\bigl(\ln\bar\phi+\tfrac12\sigma^2_{\ln\bar\phi}\bigr)^2}{2\sigma^2_{\ln\bar\phi}}\right],
\label{eq:superstat}
\end{equation}
with the signs of Eq.~\eqref{eq:watson}. It has no closed form, the Laplace transform of a lognormal not being elementary, and is evaluated by quadrature over $\bar\phi$.

The superposition does not change the leading aligned mean-square tilt. For a girdle
\begin{equation}
\langle\mu^2\rangle\simeq\frac{1}{2b_0},
\label{eq:m2int}
\end{equation}
and $\langle1-\mu^2\rangle\simeq1/b_0$ for a bipolar state, with no dependence on $\sigma^2_{\ln\bar\phi}$: the aligned conditional moment is linear in $\bar\phi$ and $\langle\bar\phi\rangle=1$. At finite $b_0$ the conditional moments are no longer linear in $\bar\phi$ and the tilt statistics shift; at $b_0=6$ and $\sigma^2_{\ln\bar\phi}=1.2$, $\langle\mu^2\rangle$ falls by $18\%$ for a column and $\langle1-\mu^2\rangle$ by $19\%$ for a plate, and the angle moments shift by different amounts.

\begin{figure}
\includegraphics[width=\textwidth]{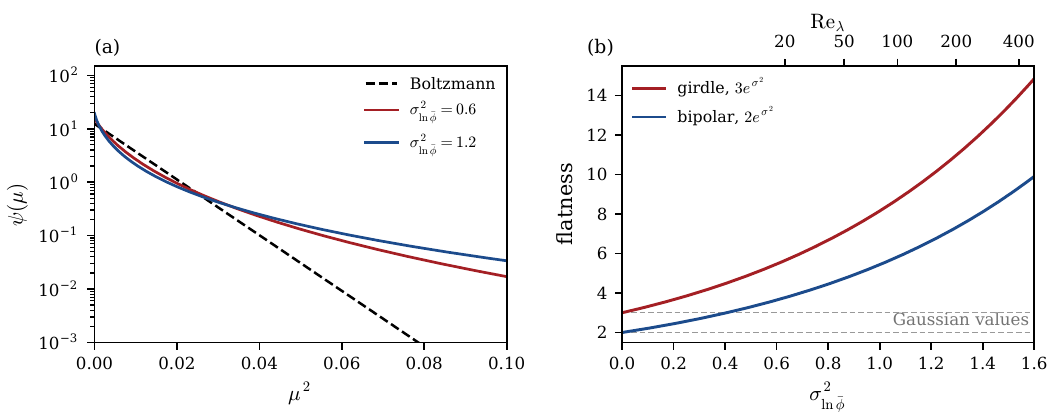}
\caption{Dissipation intermittency in the aligned limit. (a) The superposition \eqref{eq:superstat} for a girdle at $b_0=120$ with two log-dissipation variances (solid), against the Boltzmann density of the same $b_0$ (dashed). (b) The flatness \eqref{eq:flatness} and its bipolar counterpart \eqref{eq:flatness2}, with the Taylor-scale Reynolds number from Eq.~\eqref{eq:sigmaRe} on the upper axis.}
\label{fig:intermittency}
\end{figure}

The mixture is a superposition of exponentials in $\mu^2$ with positive weights, so $\ln\psi$ is convex in $\mu^2$, where a single Boltzmann density gives a straight line; upward curvature is therefore consistent with a mixture of concentrations, though it does not establish one, still less a lognormal one. The aligned-limit moments of the mixture have closed forms. With $\langle\cdot\rangle_b$ an average over a single density of concentration $b$, from $\langle\mu^{2n}\rangle_b\to(2n-1)!!\,(2b)^{-n}$ in the aligned limit and $\langle\bar\phi^{\,n}\rangle=\exp[\tfrac12n(n-1)\sigma^2_{\ln\bar\phi}]$,
\begin{equation}
\langle\mu^{2n}\rangle\simeq\frac{(2n-1)!!}{(2b_0)^{n}}\,
\exp\!\Bigl[\tfrac12 n(n-1)\,\sigma^2_{\ln\bar\phi}\Bigr],
\label{eq:ssmoments}
\end{equation}
and the first new information is at $n=2$, where the flatness takes the closed form
\begin{equation}
\mathcal{F}=\frac{\langle\mu^4\rangle}{\langle\mu^2\rangle^2}\simeq3\exp\bigl(\sigma^2_{\ln\bar\phi}\bigr),
\label{eq:flatness}
\end{equation}
Gaussian as $\sigma^2_{\ln\bar\phi}\to0$ and, with the refined-similarity form $\sigma^2_{\ln\bar\phi}\simeq\zeta\ln\mathrm{Re}_\lambda+\mathrm{const}$ for a constant $\zeta$, growing as a power of $\mathrm{Re}_\lambda$, like velocity-derivative flatness \citep{Sreenivasan1997} (figure~\ref{fig:intermittency}b). Higher moments need stronger alignment for these asymptotic forms to hold, since they weight rare intense events more heavily. The most visible effect is in the tails (figure~\ref{fig:intermittency}): at $\sigma^2_{\ln\bar\phi}=0.6$ the mixture exceeds the Boltzmann density of the same $b_0$ by two orders of magnitude at $\mu^2=0.1$, so large tilts, produced by rare intense dissipation events, are far more frequent than under Gaussian forcing. The peak of the girdle rises as well, by $\langle\bar\phi^{-1/2}\rangle=\exp(3\sigma^2_{\ln\bar\phi}/8)$, at a fixed leading mean-square tilt, while the mean of $|\mu|$ falls by the factor $\exp(-\sigma^2_{\ln\bar\phi}/8)$, by $7\%$ and $14\%$ at $\sigma^2_{\ln\bar\phi}=0.6$ and $1.2$. \cite{Pumir2026} found a qualitatively similar sharpening when they made their stochastic model intermittent, their distributions becoming more peaked as $\mathrm{Re}_\lambda$ grew while the mean absolute orientation cosine changed little. These leading-order moments also apply to other systems with locally quadratic restoring potentials, provided the conditional fluctuations are Gaussian and their variance follows the same slowly varying lognormal modulation.

The same factorisation gives in closed form the constants $C_n(\mathrm{Re}_\lambda)=\int_0^\infty\gamma^{2n}P_\gamma\,\dd\gamma$, with $P_\gamma$ the density of the shear rate folded onto $\gamma\ge0$, that govern the rapid-settling orientation moments in the map construction of \cite{Roy2023}, previously obtained by quadrature. With the shear rate, in units of its root-mean-square value, written as $\gamma=h\bar\phi^{1/2}$, where $h$ is a unit-variance Gaussian independent of $\bar\phi$ \citep{Koch2002},
\begin{equation}
C_n(\mathrm{Re}_\lambda)=(2n-1)!!\,\exp\bigl[\tfrac12n(n-1)\sigma^2_{\ln\bar\phi}\bigr],
\label{eq:Cn}
\end{equation}
so that
\begin{equation}
C_4/C_2^{\,2}=\tfrac{35}{3}\exp\bigl(4\sigma^2_{\ln\bar\phi}\bigr),
\label{eq:kurt}
\end{equation}
a ratio of eighth to squared fourth moments of $\gamma$, independent of the mean of $\ln\bar\phi$. The number of restored directions (Sec.~\ref{sec:topology}) also sets the Gaussian baseline of the flatness. Equation~\eqref{eq:flatness} carries the value three because a girdle restores one tangent direction; for a bipolar state $1-\mu^2$ is a sum of two squared Gaussians, and
\begin{equation}
\mathcal{F}_{\rm bipolar}=\frac{\langle(1-\mu^2)^2\rangle}{\langle1-\mu^2\rangle^2}\simeq2\exp\bigl(\sigma^2_{\ln\bar\phi}\bigr),
\label{eq:flatness2}
\end{equation}
so a measured flatness must be normalised by two or by three according to the attracting set.

At the slip maximum, where the closure concentration is $6.0$ for the column and $7.5$ for the plate, the mixture with $\sigma^2_{\ln\bar\phi}=1.2$ reduces $\mathcal S$ by $25\%$ and $30\%$ if the correlation time does not follow the dissipation, so the rms slip falls by $14\%$ and $16\%$. If instead $\tau_c$ follows the local Kolmogorov time, the local concentration follows from Eq.~\eqref{eq:linresp} with that time rather than from $b_0/\bar\phi$, and $\mathcal S$ falls by $23\%$ and $26\%$. The maximum moves too: $c^*$ rises from $0.925$ to $1.031$ for the column and from $0.998$ to $1.141$ for the plate, or to $0.941$ and $1.045$ when $\tau_c$ follows the dissipation. Read with the constant of a single Boltzmann density, a mixture peak would underestimate $F$ by up to $20\%$ and $23\%$ and, with $F$ known, overestimate $\varepsilon$ by up to $55\%$ and $71\%$.

The mixture also assumes that the dissipation stays fixed while the orientation equilibrates. Multiplying the forcing by a lognormal amplitude $\bar\phi^{1/2}(t)$ of correlation time $\tau_{\bar\phi}$ (Appendix~\ref{app:numerics}) tests this assumption: for the $\kappa=10$ column at $\lambda_r\tau_c=3$ and $\sigma^2_{\ln\bar\phi}=1.2$, the measured $\langle\mu^2\rangle$ is $0.83$, $0.87$, $0.90$ and $0.90$ times the static superposition of closure densities for $\tau_{\bar\phi}/\tau_K=0.3$, $1$, $4$ and $20$, and $0.91$ when each particle keeps a frozen amplitude. The modulated runs level off by $\tau_{\bar\phi}\approx4\tau_K$, about $1\%$ below the frozen-amplitude value, and the remaining $9$--$10\%$ is the closure's own error, about twice its value of $5\%$ at uniform dissipation (Sec.~\ref{sec:watson}), because the mixture weights the weakly aligned states in which the closure is least accurate. Through the crossover the mixture is thus a sensitivity estimate, and it indicates that the single density overestimates the lateral slip; anything weighted towards large tilts, such as the fourth moment, the flatness or the probability of finding a crystal far from broadside-on, needs an account of the non-Gaussian forcing, of which the mixture is one tractable model.

\section{Tests against simulated turbulence}
\label{sec:tests}

This section examines the forcing felt along a trajectory, which the closure takes from fixed-orientation statistics, and compares the results with direct numerical simulation.

\subsection{Correlation between orientation and the sampled gradients}
\label{sec:conditioning}

The closure cannot absorb a correlation between $\p$ and the instantaneous gradient, which Sec.~\ref{sec:forcing} assumes absent. Gravity alone does not spoil the one-time statistics: at fixed $\p$ the velocity $\mathbf u(\mathbf x,t)+\w(\p)$ sampled by an inertialess particle is divergence free, which removes the usual mechanism of preferential concentration (though it does not prove uniform sampling, $\p$ and $\mathbf x$ evolving together), and settling changes the time correlation, through the crossing-trajectory effect \citep{Csanady1963} represented by the time $\ell/w$. The orientation is, however, correlated with the gradient, because the orientation a particle has reached depends on the forcing that brought it there. For forcing continuous in time, as here, the conditional mean $\langle\mathbf f\,|\,\p\rangle$ is an ordinary conditional expectation and does not vanish: in a stationary state without probability current it cancels the deterministic drift (Appendix~\ref{app:cond}). Ideal white noise has no instantaneous value, and the corresponding statement is the zero-current balance of drift and diffusive flux from which the Boltzmann density follows. Neither is a drift to be added to a Fokker--Planck equation, and neither bears on the accuracy of the closure, which depends on the strength and distribution of the sampled forcing.

We measured the strength and distribution of the sampled forcing in both synthetic models. Under the projected-vector forcing, along the trajectories of Appendix~\ref{app:numerics} for the $\kappa=10$ column, $\langle|\mathbf f|^2\rangle=0.231$ at $\Sv=1.78$ and $0.218$ at $\Sv=0.5$, against the fixed-orientation value $2\sigma_f^2=0.263$: the forcing felt is $12\%$ and $17\%$ weaker. At the attained orientation $\langle(\bm\xi\cdot\p)^2\rangle=0.163$ and $0.176$, against the isotropic $\sigma_f^2=0.131$, so the ambient vector points along $\p$ more often than chance, having driven the particle there, and the tangential part that survives the projection is depleted. That mechanism is specific to the vector model, whose forcing is even under $\p\to-\p$ in each realisation (Sec.~\ref{sec:watson}). The weakening is not: in the tensor-forced runs of table~\ref{tab:cond}, at $F\Sv^2=1.0$, the conditional second moments weighted by the population of their bins give $\langle|\mathbf f|^2\rangle=0.233$ for the column and $0.247$ for the plate, $11\%$ and $7\%$ below the fixed-orientation values $0.263$ and $0.266$. The conditional second moments also depend on orientation and differ between the meridional and azimuthal directions (Appendix~\ref{app:cond}).

We do not turn these measurements into a corrected equation for the orientation alone. For Gaussian gradients such an equation could start from the identity of \cite{Novikov1965}, which writes averages of $\mathbf f(\p(t),t)$ through the gradient covariance and the response of $\p$ to the gradient history; that response decays as $e^{-\lambda_r s}$ only in the aligned linear limit and only in the restored directions, and not at all for the azimuth of a column. A Gaussian calculation would also miss the source of the rod--vorticity alignment, which arises because $\p$ and $\bm\omega$ obey similar equations of motion \citep{Pumir2011}, whereas the vorticity of a prescribed Gaussian gradient is not stretched by its strain. The net effect of the correlation on the concentration can be judged only against direct simulation, with which the next section compares the results for small plates and columnar crystals.

\subsection{Comparison with direct numerical simulation}
\label{sec:anand}

\subsubsection{The orientation density}

\cite{Anand2020} report orientation distributions of $m=|\p\cdot\ghat|$ for oblate spheroids settling in homogeneous isotropic turbulence at $\mathrm{Re}_\lambda=150$, for three aspect ratios at fixed settling Reynolds number, $\mathrm{Re}_s=0.8$. Their model shares our starting point: the gradient covariances they quote are Eqs.~\eqref{eq:Ecov}--\eqref{eq:Wcov}, their rapid-settling variance carries the combination $5+3\mathcal{B}^2$ of Eq.~\eqref{eq:fvar}, and their gravity-induced torque is evaluated with the quasi-static slip, the mobility acting on gravity, as here. Their particles have translational and rotational inertia, at Stokes numbers $\mathrm{St}_\eta=0.24$--$0.40$, but the slip fluctuations that inertia produces do not enter their torque, so the comparison tests the overdamped torque balance in resolved turbulence rather than the finite-inertia mechanism of \cite{Gustavsson2021}. Rotational inertia is not negligible, however: for a thin disc the rotational response time is $\tau_r\simeq0.3\tau_p$, with $\tau_p$ the particle response time of their model, so $\lambda_r\tau_r\approx0.1$--$0.5$. Under Ornstein--Uhlenbeck forcing the linearised oscillator of Sec.~\ref{sec:gradflow} has a tilt variance $1+\lambda_r\tau_r/(1+\lambda_r\tau_c+\tau_r/\tau_c)$ times the overdamped value \eqref{eq:linresp}, so with $\tau_c=\tau_K$ rotational inertia would raise the linear tilt variance of these runs by $4$--$10\%$.

We digitise their Fig.~1(a) over $0\le m\le0.99$, three to six decades in density, and plot it against $m^2$, on which a single Boltzmann density is a straight line (figure~\ref{fig:anand}a). A single Boltzmann density does not describe these data: its residuals are large and systematic, the measured density curving upward in $m^2$ and rising towards the broadside orientation faster than a single exponential. Section~\ref{sec:intermittency} predicts curvature of this sign. Mixing Boltzmann densities over a lognormal dissipation with the variance estimated from the Reynolds number, $\sigma^2_{\ln\bar\phi}=1.20$, and fitting only $b_0$ cuts the residuals five- to sevenfold with the same number of parameters; letting the variance float lowers them further and returns $\sigma^2_{\ln\bar\phi}=0.84$, $0.87$ and $0.90$, three values that agree with one another and lie $25$--$30\%$ below Eq.~\eqref{eq:sigmaRe}. The fitted variance is an effective parameter of the model rather than a measurement of the dissipation statistics: it absorbs the torque approximation, the temporal response, the sampling along trajectories and the digitisation, and for $\kappa=0.05$ it rises from $0.90$ to $0.97$ when the points above $m=0.95$ are dropped, those of the two thinner discs moving by $0.02$ or less. The bin convention of the source matters near the pole. Its markers are spaced by $0.02$ in $m$; read as means over bins of that width, the last $\kappa=0.05$ marker, $52.3$ at $m=0.99$, would carry probability $1.05$ by itself and that curve $1.12$ in all, against $1.01$ and $0.96$ for the other two. Averaging the model over bins of width $0.02$ lowers the one-parameter residual of the mixture for $\kappa=0.05$ from $0.26$ to $0.16$ and raises the variance fitted with $b_0$ to $0.97$, with residual $0.084$, and that of the thinner discs to $0.85$ and $0.89$; the preference for the mixture is unchanged. The fits constrain only the unconditional density, so they cannot establish that the density at a given local dissipation is a Boltzmann density.

The concentration can also be predicted from the run parameters. The supplementary material of \cite{Anand2020} lists the Froude number $\mathrm{Fr}_\eta=U_s/u_\eta$ of each run, $2.1$, $2.7$ and $3.7$ for $\kappa=0.01$, $0.02$ and $0.05$. Their $U_s=\tau_pg$ is the slip with the symmetry axis vertical, the broadside orientation of a plate, so $\Sv_b=\mathrm{Fr}_\eta$; and their rapid-settling variance (their equation 5) is the aligned limit $\langle1-\mu^2\rangle=1/b$ of Eq.~\eqref{eq:b}, with their resistance functions combining into $F_b=0.329$, $0.325$ and $0.311$ (Appendix~\ref{app:mobility}). Equation~\eqref{eq:b} with $\tau_c=\tau_K$ then gives $b=13.4$, $29.9$ and $84.3$, with nothing adjusted. The mixture fits return $b_0$ about a quarter above these at $\sigma^2_{\ln\bar\phi}=1.20$, and within $11\%$ of them when the variance is fitted as well. With nothing fitted at all, the mixture at the predicted concentration follows the data to within a typical factor $\exp(\mathrm{rms})$ of $1.5$ to $2.0$ (table~\ref{tab:anandfits}) over three to six decades, whereas a single Boltzmann density at the same concentration falls short at large tilt by two to thirty decades. The densities are digitised, and their normalisation on $0\le m\le1$ is assumed.

\subsubsection{Moments through the crossover}
\label{sec:anandmoments}

These densities come from three runs at one settling Reynolds number. The main panels of Fig.~2 of \cite{Anand2020} give, for every run, the measured moments $X=\langle1-m^2\rangle$, the spread about the broadside orientation, and $Y=\langle(1-m)^2\rangle$, which is of fourth order in the tilt near alignment and weights large tilts more strongly; at $\mathrm{Re}_\lambda=150$ the runs span the crossover, from $F_b\Sv_b^2=0.10$ to $4.2$. We read seven of the nine runs (two overlap near $\mathrm{Fr}_\eta=2.7$; Appendix~\ref{app:numerics}) and compare them in figure~\ref{fig:anand}(c,d) with the single Boltzmann density and the mixture at $\sigma^2_{\ln\bar\phi}=1.20$, both at the concentration \eqref{eq:b} predicted from the run parameters, nothing fitted; the curves are drawn for $\kappa=0.02$, those for $\kappa=0.01$ and $0.05$ differing by less than $1\%$. The spread is predicted: from $F_b\Sv_b^2=0.8$ upwards the single density is within $3$--$13\%$ of the measured $X$ and the mixture within $0.5$--$11\%$. The tails are not, or not uniformly: from $F_b\Sv_b^2=1.3$ upwards the single density underestimates $Y$ by factors of $2.4$--$2.9$, and the mixture overestimates it by an amount that grows with settling, from $16\%$ to $61\%$. At the three weakest torques, $F_b\Sv_b^2=0.10$--$0.26$, the data lie between the two forms, their departure from the uniform value $X=2/3$ being $1.6$--$1.9$ times that of the single density, a deficit of the sign and size found against the stochastic simulations (Sec.~\ref{sec:watson}). Reading the markers at their pixel centroids, or the settling numbers from the rounded values of the supplementary table, changes these ranges (Appendix~\ref{app:numerics}) but not their ordering: the single density stays low by factors of $2.4$--$3.1$ in $Y$, and the excess of the mixture grows with settling. The runs have Stokes numbers $0.06$--$0.40$, so the comparison is not strictly overdamped. Agreement in $X$ supports the predicted width through the crossover, but it cannot identify the dissipation mechanism, and it does not locate the slip peak, which depends on the fourth moment through $\mathcal S$ (Sec.~\ref{sec:peak}). A corresponding test for columns, and a measurement of the slip peak in resolved turbulence, remain to be made.

\subsubsection{The fourth moment}

\begin{figure}
\includegraphics[width=\textwidth]{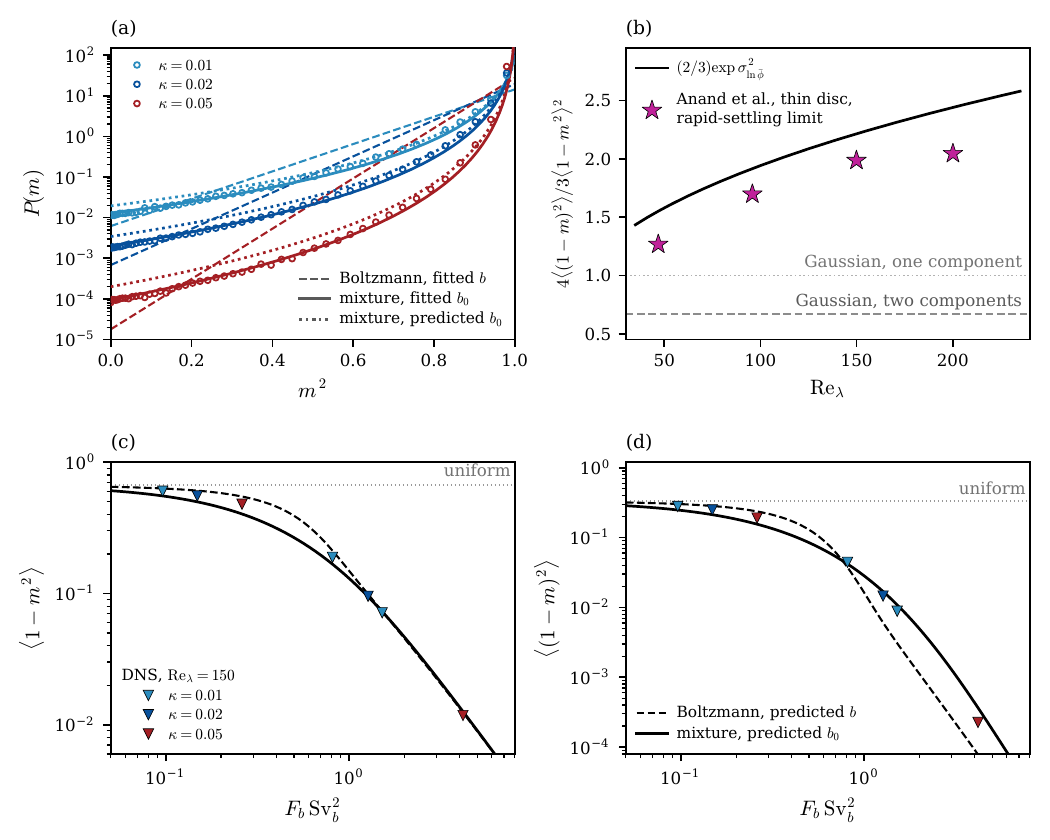}
\caption{Comparison with the plate DNS of \cite{Anand2020} at $\mathrm{Re}_\lambda=150$. (a) Densities of $m=|\p\cdot\ghat|$ at $\mathrm{Re}_s=0.8$: symbols, the data; dashed, a single Boltzmann density with $b$ fitted; solid, the superposition \eqref{eq:superstat} with $\sigma^2_{\ln\bar\phi}=1.20$ and $b_0$ fitted; dotted, the superposition at the predicted concentration \eqref{eq:b}. (b) Their gradient-based ratio $Q$ against $(2/3)\exp(\sigma^2_{\ln\bar\phi})$, Eq.~\eqref{eq:flatness2}. (c), (d) Measured moments (triangles) against $F_b\Sv_b^2$, with the single density (dashed) and the superposition (solid) at the predicted concentration; dotted, the uniform density.}
\label{fig:anand}
\end{figure}

A third test, with no fitted parameter, uses the fourth moment; it tests the amplitude model of Sec.~\ref{sec:intermittency} rather than the three densities of figure~\ref{fig:anand}(a). \cite{Anand2020} evaluate $Q=4\langle(1-m)^2\rangle/3\langle1-m^2\rangle^2$ (with the signed $\mu$, $\langle(1-\mu)^2\rangle$ would tend to $2$, not $0$, for a plate equally likely at either pole) for a thin disc in the rapid-settling limit, where the tilt follows the instantaneous gradient, from their theory with fourth moments of the velocity gradients measured in their direct numerical simulations (DNS; inset of their Fig.~2(b)), and trace its growth with $\mathrm{Re}_\lambda$ to the intermittency of the gradients. In that limit their expression reduces to $Q=\tfrac{25}{2}G_2$, with $G_2=\langle(\tau_K\,\partial u_1/\partial x_2)^4\rangle$, so $Q$ is $2/9$ of the flatness of a transverse velocity gradient: a property of the gradient statistics, not a fourth moment of the measured densities. For small tilt $Q=\langle\varphi^4\rangle/3\langle\varphi^2\rangle^2$, and the superposition gives it the closed form $\tfrac23\exp(\sigma^2_{\ln\bar\phi})$ through Eq.~\eqref{eq:flatness2}. The Gaussian value is $2/3$ because a disc is restored in two tangent directions, and $G_2=4/75$ for Gaussian gradients gives the same value; the unity that \cite{Anand2020} quote as Gaussian holds for a single tilt component. With $\sigma^2_{\ln\bar\phi}$ from Eq.~\eqref{eq:sigmaRe} (figure~\ref{fig:anand}b) the prediction runs $12$--$22\%$ above their four values, $1.27$, $1.70$, $1.99$ and $2.04$ at $\mathrm{Re}_\lambda=47$, $96$, $150$ and $200$, but reproduces their level and trend: the log-dissipation variance implied by $Q$ grows as $\dd\sigma^2_{\ln\bar\phi}/\dd\ln\mathrm{Re}_\lambda\approx0.34$, against $0.31$ for the parametrisation, a consistency check of the amplitude model rather than a measurement of the exponent from four points. Every value lies far above $2/3$, and above unity as well, so normalising by three, as though the tilt had one component, understates the departure from Gaussianity. The moments of Sec.~\ref{sec:anandmoments} give $Q$ directly for the three runs with $F_b\Sv_b^2\ge1.3$: $2.14$, $2.32$ and $2.17$, close to their inset value $1.99$ at $\mathrm{Re}_\lambda=150$ and to the aligned limit $2.21$ of the superposition, whereas the superposition at the concentrations of those runs gives $2.5$--$3.1$ and the single density $0.68$--$0.83$.

\subsubsection{Finite Stokes number}

\begin{figure}
\includegraphics[width=\textwidth]{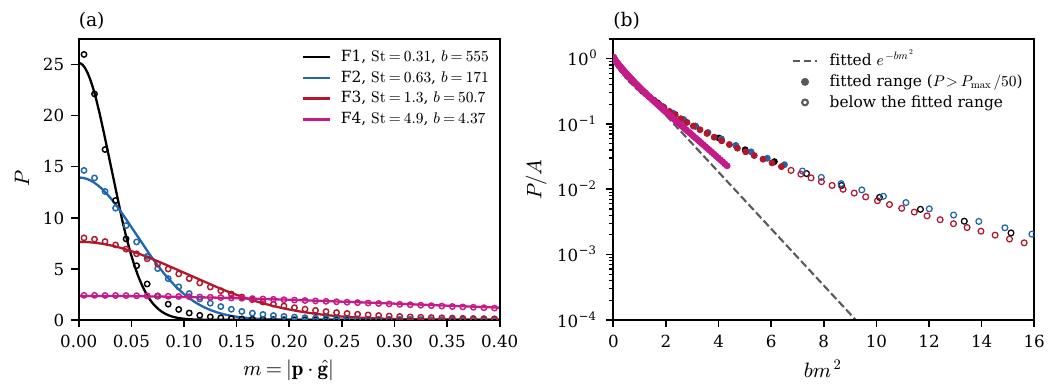}
\caption{Orientation distributions of a column of aspect ratio four from Fig.~2 of \cite{Pumir2026}, in four flows labelled by their Stokes number. (a) The data (symbols) and fitted curves $A\exp(-bm^2)$ (lines). (b) The data divided by $A$, against $bm^2$, on which each fit is the dashed line $e^{-bm^2}$; filled symbols lie in the fitted range.}
\label{fig:pumir}
\end{figure}

The runs of \cite{Pumir2026} are at finite Stokes number, outside the overdamped regime (Sec.~\ref{sec:application}), and their distributions show what particle inertia does to the shape. Their Fig.~2 gives $P(|\p\cdot\ghat|)$ for a column of aspect ratio four in four flows, at Stokes numbers from $0.31$ to $4.9$, on linear axes. We use the vertices of the published curves, read from the vector graphics of the article, which resolve the density to about $10^{-3}$, three to four decades below the peak (Appendix~\ref{app:numerics}). Fitted on the linear axes of the source to the points above a fiftieth of the peak, curves $A\exp(-bm^2)$ describe the cores to $3$--$3.5\%$ of the peak height in rms, and to $1.5\%$ in the least aligned flow, with concentrations $555$, $171$, $50.7$ and $4.37$ (figure~\ref{fig:pumir}a). The core concentration depends on the extraction and on the fitted range: the same fit to an independent digitisation of the plotted curves gave $667$, $187$, $52$ and $4.4$, and raising the threshold to half the peak gives $741$, $222$, $64$ and $4.9$. Divided by the fitted amplitude and plotted against $bm^2$ (figure~\ref{fig:pumir}b), the three aligned flows approximately collapse onto a common profile, which curves upward on these semilogarithmic axes where a single Boltzmann density is straight. At these Stokes numbers that curvature need not come from intermittency (Sec.~\ref{sec:intermittency}) alone. Within the fitted range the data rise above the fit by factors up to $12$--$16$, and by $150$--$190$ at $bm^2=10$. The normalised Boltzmann density at the fitted concentration underestimates the tabulated mean-square tilt by $36$--$38\%$ in the three aligned flows. The fits therefore describe the width of the core, not the shape of the distribution, and give no evidence that the Boltzmann form survives inertia. The fitted concentrations describe the simulated distributions; they are not predictions. The prediction is the Stokes-drag overdamped reference of Sec.~\ref{sec:application}, computed from the run parameters with nothing fitted, and the three quantities differ: in flow F1, the least turbulent, the tabulated rms tilt is $2.15^\circ$, the Boltzmann density at the fitted core concentration gives $1.72^\circ$, and the overdamped reference gives $0.224^\circ$. Over the four flows the core concentrations are $1.7$--$2.9\%$ of Eq.~\eqref{eq:b}, a core six to eight times broader in angle, and the tabulated rms tilt is $6.6$--$9.6$ times the reference, which indicates what inertia adds beyond the closure; the finite-$\Rep$ drag left out of the reference would raise it by about $42\%$ for these particles.

\begin{table}
\caption{\label{tab:anandfits}Prediction and fits for the orientation distributions of \cite{Anand2020} in figure~\ref{fig:anand}(a), digitised over $0\le m\le0.99$ and grouped by the number of fitted parameters. Every density is normalised on $0\le m\le1$, with no free amplitude, and evaluated at the abscissae of the markers (Sec.~\ref{sec:anand} gives the effect of averaging over bins). Residuals are rms deviations of $\ln P$ over the digitised points, so $\exp(\mathrm{rms})$ is the typical factor between model and data. $b_{\rm pred}$ is Eq.~\eqref{eq:b} with $\Sv_b=\mathrm{Fr}_\eta$ and $F_b$ from Appendix~\ref{app:mobility}. The mixture is Eq.~\eqref{eq:superstat} with $\sigma^2\equiv\sigma^2_{\ln\bar\phi}=1.20$ from Eq.~\eqref{eq:sigmaRe} unless $\sigma^2$ is fitted; the power-law form is Eq.~\eqref{eq:powerpdf}, bipolar, with $p=3.43$ from Eq.~\eqref{eq:powerpar} unless $p$ is fitted, and the $\sigma^2$ implied by a fitted $p$ follows by inverting Eq.~\eqref{eq:powerpar}.}
\centering\small
\begin{tabular*}{\textwidth}{@{\extracolsep{\fill}}lccc}
\hline
 & $\kappa=0.01$ & $\kappa=0.02$ & $\kappa=0.05$\\
\hline
$\Sv_b=\mathrm{Fr}_\eta$ & $2.1$ & $2.7$ & $3.7$\\
$b_{\rm pred}$ & $13.4$ & $29.9$ & $84.3$\\[3pt]
\multicolumn{4}{l}{\emph{rms residual, nothing fitted}}\\
\quad Boltzmann, $b=b_{\rm pred}$ & $3.7$ & $13.4$ & $49.5$\\
\quad mixture, $b_0=b_{\rm pred}$ & $0.38$ & $0.49$ & $0.71$\\
\quad power law, $\alpha=b_{\rm pred}/(p-2)$ & $0.17$ & $0.21$ & $0.69$\\[3pt]
\multicolumn{4}{l}{\emph{rms residual, one parameter fitted}}\\
\quad Boltzmann, $b$ & $0.71$ & $1.14$ & $1.92$\\
\quad mixture, $b_0$ & $0.13$ & $0.16$ & $0.26$\\
\quad power law, $\alpha$ & $0.073$ & $0.14$ & $0.43$\\[3pt]
\multicolumn{4}{l}{\emph{rms residual, two parameters fitted}}\\
\quad mixture, $b_0$ and $\sigma^2$ & $0.052$ & $0.058$ & $0.18$\\
\quad power law, $\alpha$ and $p$ & $0.072$ & $0.050$ & $0.11$\\[3pt]
\multicolumn{4}{l}{\emph{fitted parameters}}\\
\quad Boltzmann, $b$ & $7.7$ & $10.3$ & $14.2$\\
\quad mixture, $b_0$ at $\sigma^2=1.20$ & $16.6$ & $37.2$ & $107$\\
\quad mixture, $\sigma^2$ and $b_0$ & $0.84$, $14.8$ & $0.87$, $30.6$ & $0.90$, $79.7$\\
\quad power law, $p$ (implied $\sigma^2$) & $3.51$ ($1.08$) & $3.90$ ($0.75$) & $4.42$ ($0.53$)\\
\hline
\end{tabular*}
\end{table}

\subsection{A form for comparison with data}
\label{sec:fitform}

The comparisons above use the mixture \eqref{eq:superstat} directly: a one-dimensional quadrature with one free parameter, $b_0$, once $\sigma^2_{\ln\bar\phi}$ is estimated from Eq.~\eqref{eq:sigmaRe}. Where a closed-form density is wanted, as in a scattering calculation, a power-law density, normalised on $[0,1]$, can stand in for it:
\begin{equation}
\psi(\mu)=\frac{\bigl(1+\alpha\mu^2\bigr)^{-p}}{{}_2F_1\bigl(p,\tfrac12;\tfrac32;-\alpha\bigr)}
\quad\text{(girdle)},
\qquad
\psi(\mu)=\frac{\bigl[1+\alpha(1-\mu^2)\bigr]^{-p}}
{(1+\alpha)^{-p}\,{}_2F_1\bigl(p,\tfrac12;\tfrac32;\tfrac{\alpha}{1+\alpha}\bigr)}
\quad\text{(bipolar)},
\label{eq:powerpdf}
\end{equation}
where $\alpha>0$ is a fitting scale multiplying $\mu^2$ for a girdle and $1-\mu^2$ for a bipolar state; the Boltzmann form is recovered as $p\to\infty$ at fixed $p\alpha$. Matching the aligned-limit variance \eqref{eq:m2int} and flatness \eqref{eq:flatness} of the mixture, or for a bipolar state $\langle1-\mu^2\rangle=1/b_0$ and Eq.~\eqref{eq:flatness2}, fixes both parameters in terms of $b_0$ and $E\equiv\exp\sigma^2_{\ln\bar\phi}$:
\begin{equation}
p=\frac{5E-3}{2(E-1)},\quad \alpha=\frac{b_0}{p-3/2}\ \ \text{(girdle)};
\qquad
p=\frac{3E-2}{E-1},\quad \alpha=\frac{b_0}{p-2}\ \ \text{(bipolar)} .
\label{eq:powerpar}
\end{equation}
The exponent falls towards $5/2$ and $3$ as $E$ grows, so the tail is a power law of low index in the tilt. At $\mathrm{Re}_\lambda=150$, where $E=3.32$, Eq.~\eqref{eq:powerpar} gives $p=3.43$ for a bipolar state, just below the $3.51$, $3.90$ and $4.42$ fitted, together with $\alpha$, to the three distributions of figure~\ref{fig:anand}(a).

Table~\ref{tab:anandfits} compares the two forms at equal numbers of fitted parameters: with none fitted the surrogate does better, and with one or two neither is consistently better, as expected of a form that carries the mixture's aligned-limit variance and flatness. The forms differ in the meaning of their second parameter. In the mixture it is an effective log-dissipation variance, and fitting it returns nearly one value, $0.84$ to $0.90$, for all three shapes, as expected of a property of the flow; in the surrogate it is an exponent, tied to that variance only through the aligned-limit moments, and the variances implied by its fitted exponents drift with shape, from $1.08$ to $0.53$. We would use Eq.~\eqref{eq:powerpdf} where a closed-form density is wanted and Eq.~\eqref{eq:superstat} where an effective dissipation variance is to be fitted or carried from one flow to another.

\section{Application to atmospheric ice and frazil}
\label{sec:application}

The crossover $F\Sv^2\sim1$ is the regime of practical interest: for crystals of tens of microns in cloud turbulence, and for frazil discs in supercooled water, the alignment parameter passes through order unity within a single cloud or plume. The formulas apply as they stand only if the translational and rotational response times of the particle are short against the forcing and alignment times, $\max(\tau_t,\tau_r)\ll\min(\tau_K,\tau_c,\lambda_r^{-1})$, so a small Stokes number based on $\tau_K$ alone is not enough when the alignment is fast. In air this is restrictive: ice is about a thousand times denser than air, and \cite{Gustavsson2021} showed that translational inertia then feeds slip-velocity fluctuations into the angular equation and raises the mean-square tilt by orders of magnitude for slender columns and thin plates, even at small Stokes number, finding overdamped dynamics for platelets in cloud conditions only for $a\sqrt\kappa\ll18$\,\si{\micro\metre}. Outside that window the present results are the overdamped reference, the alignment a crystal would reach if its slip did not fluctuate, and not a prediction of the observed distribution: at finite inertia $\w$ is itself a fluctuating vector coupled to the orientation and to the sampled flow, and eliminating it can produce multiplicative forcing and non-gradient drift, so whether the Boltzmann form survives there is an open question.

The overdamped condition is easier to meet in water. A frazil disc $1\,$mm across and $50$\,\si{\micro\metre} thick ($\kappa=0.05$) rises broadside-on at about $1.5\,\mathrm{mm\,s^{-1}}$ at \SI{0}{\celsius}, with a diameter-based Reynolds number near $0.8$. In turbulence with $\varepsilon=10^{-7}$--$10^{-6}\,\mathrm{m^2\,s^{-3}}$ its diameter is $0.36$--$0.65$ Kolmogorov lengths, its Stokes number is of order $10^{-3}$, and $\Sv_b=2.3$--$1.3$, so with $F_b\simeq0.31$ it sits in the crossover, at $b\simeq16$--$3$. For a nearly neutrally buoyant disc the added-mass and history forces need assessment before quantitative use; the particles of the settling experiments of \cite{Lopez2017} and \cite{Roy2019} are in a similar position.

A scattering measurement returns the orientation average of a cross-section. In the Rayleigh regime of centimetre- and millimetre-wavelength radar the polarisability of a spheroid is $\bm{\alpha}=\alpha_\parallel\p\p+\alpha_\perp(\I-\p\p)$, so the co-polar return at horizontal polarisation $\hat{\mathbf h}$ is governed by $(\p\cdot\hat{\mathbf h})^2$ and its square, which for a density axisymmetric about gravity average to
\begin{equation}
\bigl\langle(\p\cdot\hat{\mathbf h})^2\bigr\rangle=\tfrac12\bigl(1-\langle\mu^2\rangle\bigr),\qquad
\bigl\langle(\p\cdot\hat{\mathbf h})^4\bigr\rangle=\tfrac38\bigl(1-2\langle\mu^2\rangle+\langle\mu^4\rangle\bigr).
\label{eq:radar}
\end{equation}
The orientation factors in reflectivity and differential reflectivity are thus set by $\langle\mu^2\rangle$ and $\langle\mu^4\rangle$, which follow from the recursion \eqref{eq:momrec} for a Boltzmann density and with one further average over $\bar\phi$ for the mixture. Lidar needs the full density and a non-Rayleigh calculation; the sensitivity to the tilt width found by \cite{Okamoto2020} shows why the width matters, although their plates, larger than $200$\,\si{\micro\metre}, lie beyond the overdamped regime and do not test Eq.~\eqref{eq:b}. For small crystals the mean-square tilt saturates at the random-orientation values \eqref{eq:randpro} as $\Sv$ falls, where the aligned-limit law would grow without bound as $\Sv^{-4}$; that saturation is a testable signature of the crossover, and intermittency predicts tails that broaden with the Reynolds number at a fixed leading aligned variance.

Swept-volume collision kernels for crystal--droplet and crystal--crystal encounters \citep{Naso2018,Jucha2018,Sheikh2022,Sheikh2024} need $\langle|\mu|\rangle$, $\langle\mu^2\rangle$ and the slip statistics \eqref{eq:vh2}, all in closed form at any $\Sv$, where the strong-alignment expansions err by tens of per cent at the operating point (Sec.~\ref{sec:moments}); crystal--crystal kernels also need pair statistics. In laboratory settling experiments \citep{Lopez2017}, where the overdamped condition is under the experimenter's control, the lateral slip relative to the local fluid tests the peak relation of Sec.~\ref{sec:peak}, with the conventions of Sec.~\ref{sec:renorm} connecting it to quiescent-flow theory.

\begin{table}
\caption{\label{tab:dnsfloor}Measured rms tilt of settling ice columns in the DNS of \cite{Pumir2026} (their Table~III; flows F1--F4, particles P1 and P3) against the Stokes-drag overdamped reference of their model dynamics, with the torque factor $C_\tau=0.5$ of their equations of motion and the physical broadside settling number $\Sv_b=\hat U_\perp(a/R_{\rm eq})\,\Sv_{\rm P}$, with $\Sv_{\rm P}$ their settling number. $\mathrm{St}$ and $\Sv_{\rm P}$ are recomputed from the particle and flow parameters of their Tables~I and II, which reproduce the values printed in their Table~III to within $3\%$, and every computed column follows from them. The reference is the rms angle $\langle\varphi^2\rangle^{1/2}$, with $\langle\varphi^2\rangle=Z_-^{-1}\int_0^1(\arcsin\mu)^2e^{-b\mu^2}\dd\mu$ at the concentration \eqref{eq:b}, evaluated by quadrature rather than from the aligned asymptote, since run F4/P1 sits at $C_\tau F_b\,\Sv_b^2=2.5$. The finite-$\Rep$ drag of their model, not included, would raise the reference by about $24\%$ for P1 and $59\%$ for P3 in the aligned limit.}

\centering\small
\begin{tabular*}{\textwidth}{@{\extracolsep{\fill}}lccccccc}
\hline
run & $\kappa$ & St & $\Sv_{\rm P}$ & $\Sv_b$ & DNS ($^\circ$) & reference ($^\circ$) & ratio\\
\hline
F1/P1 & 3 & 0.185 & 39.0 & 22.6 & 1.77 & 0.477 & 3.71\\
F1/P3 & 5 & 0.439 & 92.8 & 39.1 & 2.31 & 0.145 & 16.0\\
F2/P1 & 3 & 0.370 & 27.6 & 16.0 & 3.47 & 0.945 & 3.67\\
F3/P1 & 3 & 0.739 & 19.5 & 11.3 & 6.40 & 1.84 & 3.47\\
F3/P3 & 5 & 1.76 & 46.4 & 19.6 & 7.71 & 0.573 & 13.5\\
F4/P1 & 3 & 2.94 & 9.79 & 5.66 & 21.2 & 6.54 & 3.24\\
F4/P3 & 5 & 6.98 & 23.3 & 9.81 & 22.5 & 2.19 & 10.3\\
\hline
\end{tabular*}

\end{table}

Comparing the overdamped reference with recent direct simulations first requires reconciling conventions. \cite{Pumir2026} tabulate the rms tilt of columns with $\kappa=3$--$5$ and semi-major axes $100$--$200$\,\si{\micro\metre} in four flows spanning $\varepsilon=4.4$--$1120\,\mathrm{cm^2\,s^{-3}}$. Their equations of motion carry a calibrated inertial-torque factor $C_\tau=0.5$, so their overdamped limit has relaxation coefficient $C_\tau F_b$, and their settling number $\Sv_{\rm P}$, built on the response time $m/(6\pi\mu_f a)$ with $a$ the semi-minor axis, exceeds the physical broadside settling number by the factor $[\hat U_\perp\,a/R_{\rm eq}]^{-1}\simeq1.7$--$2.4$. Table~\ref{tab:dnsfloor} sets their measured tilt against the overdamped reference of their own model with Stokes drag, computed from their particle and flow parameters with no adjustable constant. For the smallest columns ($\kappa=3$, $100$\,\si{\micro\metre}) the measured rms exceeds the reference by $3.2$--$3.7$ in all four flows, at Stokes numbers up to three. For the largest ($\kappa=5$, $200$\,\si{\micro\metre}) the gap opens to $10$--$16$, and the measured tilt grows from the slow to the fast settlers of a given flow, where the overdamped scaling would have it fall, as expected when the slip fluctuations that force the orientation grow with particle size \citep{Gustavsson2021}. The reference omits two effects of opposite sign. The finite-$\Rep$ drag of their model (their equations (C3) and (C4)) slows the broadside settling, from $21.6$ to $19.1\,\mathrm{cm\,s^{-1}}$ for the smallest columns and from $37.3$ to $28.3\,\mathrm{cm\,s^{-1}}$ for the largest, and strengthens the coupling of tilt to lateral slip; in the aligned limit the two together raise the reference by about $24\%$ and $59\%$, which would bring the ratios to $2.6$--$3.0$ and $6.5$--$10$. The orientation--gradient correlation of Sec.~\ref{sec:conditioning} would lower the reference. The ratios therefore indicate the inertial inflation without bounding it; both columns, with $a\sqrt\kappa\simeq58$ and $89$\,\si{\micro\metre}, are far larger than the platelets that \cite{Gustavsson2021} find overdamped. The conditional settling speeds that \cite{Pumir2026} report increase as the axis turns towards the vertical, as the anisotropic mobility of Sec.~\ref{sec:slip} implies, with a contrast between the extreme orientations of the order of the quiescent mobility ratio in the two flows that sample near-vertical orientations; but these are laboratory-frame velocities conditioned on orientation, which do not test the instantaneous slip relation, and preferential sweeping, outside the overdamped theory, contributes substantially to the mean settling enhancement.

\section{Conclusions}
\label{sec:conclusions}

A settling spheroid is held broadside-on by the fluid-inertial torque and turned away by the turbulent velocity gradients. In the overdamped limit the torque is an exact gradient flow on the sphere of orientations, with a potential quadratic in the cosine of the tilt that an anisotropic mobility only rescales, and the gradients force the orientation with no preferred direction in the tangent plane, so isotropic Gaussian white-noise forcing gives exactly a Boltzmann density, exponential in the squared cosine of the tilt. Turbulent forcing is not white, and keeping that density with a width fixed by the linear response is a finite-memory closure, accurate to $16\%$ or better in the second moment against stochastic simulations of a column and a plate with $\tau_c=\tau_K$, although at the weakest torques tested the simulated departure from random orientation is about $1.5$ times the closure's. The width depends on the alignment parameter $F\Sv^2$, the alignment rate in Kolmogorov units, and on the shape through the forcing factor $5+3\mathcal{B}^2$.

The observables then follow from the density without further approximation. The even moments give the tilt statistics, the mean vertical slip and the orientation averages of Rayleigh scattering, which at crossover settling numbers need not be close to either limiting assumption. The rms lateral slip does not grow without limit as particles settle faster, because alignment quenches the tilt that produces it; its maximum sits at $F\Sv^2$ between $0.74$ and $1.0$ for shapes from a moderately elongated column to a thin plate, or $5\%$ and $9\%$ higher for the two design shapes in tensor-forced simulations, so a measured peak constrains the torque coefficient or, read in reverse, the Kolmogorov scales, given the forcing statistics.

Beyond the closure, the results rest on three approximations. The forcing is prescribed rather than taken from the Navier--Stokes equations, its correlation time is uncalibrated, and the orientation it produces is correlated with the gradients it samples, which weakens and redistributes the forcing felt in a way that a single coefficient does not capture. Resolved turbulence has so far tested the closure only for plates. Their measured spread follows the prediction with nothing fitted, to within $13\%$ from $F_b\Sv_b^2=0.8$ upwards, and the weak-torque deficit of the closure appears there as well. At strong torque their fourth-order tilt moment lies between the predictions of the single density, $2.4$--$2.9$ times too low, and of the superposition, whose excess grows with settling. A test for columns, and a measurement of the slip peak, remain to be made. Intermittency enters through a lognormal amplitude that varies slowly compared with the alignment. The superposition reproduces the shape of simulated densities for small plates: with nothing fitted it follows them within a typical factor of $1.5$ to $2$, and fitting its concentration returns values about a quarter above Eq.~\eqref{eq:b} with $\tau_c=\tau_K$, although such fits cannot identify the mechanism (Sec.~\ref{sec:anand}); through the crossover the superposition is a sensitivity estimate. Finally, the analysis is overdamped, whereas atmospheric crystals larger than a few tens of microns carry enough inertia to inflate the tilt several times over. Within these limits, the closed-form density and its moments can replace the usual choice between random orientation and perfect alignment in scattering and microphysics calculations.

\begin{bmhead}[Acknowledgements.]The digitisation of the published data and the Langevin simulations were carried out with Claude Opus 5 to 5.5; all scientific content, analysis and conclusions are the author's.\end{bmhead}

\begin{bmhead}[Funding.]\sloppy This work was supported by the Anusandhan National Research Foundation under project ANRF/\allowbreak ARG/\allowbreak 2025/\allowbreak 010666/\allowbreak ENS and by the ISRO--IIT Madras Space Technology Cell project SP26271464\allowbreak AMISRO008746.\end{bmhead}

\begin{bmhead}[Declaration of interests.]The author reports no conflict of interest.\end{bmhead}

\begin{bmhead}[Data availability statement.]The Python scripts that generate the figures, the tables and the numbers quoted in the text, the Langevin simulation drivers with their job lists and seeds, the simulation outputs, and the data taken from \cite{Anand2020} and \cite{Pumir2026} (digitised points, vertices of published curves and transcribed parameters) are archived at Zenodo (\url{https://doi.org/10.5281/zenodo.23922242}); the scripts and data added in this version for Secs.~\ref{sec:conditioning} and~\ref{sec:anand} are available from the author.\end{bmhead}

\appendix
\renewcommand{\theHsection}{app.\Alph{section}}
\renewcommand{\theHsubsection}{app.\Alph{section}.\arabic{subsection}}
\renewcommand{\theHequation}{app.\Alph{section}.\arabic{equation}}

\section{Ellipsoidal mobility integrals}
\label{app:mobility}

For a spheroid with semi-axes $(a,a,c)$ the Oberbeck resistance gives the settling speed along the symmetry axis as $U_\parallel=F_g(\chi_0+c^2J_\parallel)/16\pi\mu_f$ and along a transverse axis as $U_\perp=F_g(\chi_0+a^2J_\perp)/16\pi\mu_f$ under a force $F_g$, with
\begin{equation}
\chi_0=\int_0^\infty\!\frac{\dd t}{\Delta},\qquad
J_\parallel=\int_0^\infty\!\frac{\dd t}{(c^2+t)\,\Delta},\qquad J_\perp=\int_0^\infty\!\frac{\dd t}{(a^2+t)\,\Delta},
\label{eq:oberbeck}
\end{equation}
where $\Delta=(c^2+t)^{1/2}(a^2+t)$, so that, relative to the volume-equivalent sphere of radius $R_{\rm eq}=(a^2c)^{1/3}$,
\begin{equation}
\hat U_\parallel=\frac{3R_{\rm eq}}{8}\bigl(\chi_0+c^2J_\parallel\bigr),
\qquad
\hat U_\perp=\frac{3R_{\rm eq}}{8}\bigl(\chi_0+a^2J_\perp\bigr).
\label{eq:Uhat}
\end{equation}
All three integrals reduce to inverse hyperbolic (prolate) or inverse trigonometric (oblate) functions of the eccentricity, tabulated by \cite{Happel1983} and \cite{Kim1991}. For the two design shapes, $\kappa=10$ gives $\mathcal{B}=0.980$, $\hat U_\parallel=0.814$, $\hat U_\perp=0.565$ and $g_M=0.249$, and $\kappa=0.03$ gives $\mathcal{B}=-0.998$, $\hat U_\parallel=0.366$, $\hat U_\perp=0.536$ and $g_M=-0.170$. The forcing factor $5+3\mathcal{B}^2$ varies only between $5$ and $8$ over the whole range of aspect ratio, which is why the peak constant of Sec.~\ref{sec:peak} depends only weakly on shape.

\subsection*{The torque coefficient}

The bare drift coefficient of Eq.~\eqref{eq:torque} is $F_0=F_D\,\mu_f\,a_{\max}^3/C_r$, where $F_D(\kappa)>0$ is the torque amplitude in $\bm{\tau}=s_\kappa F_D\,\rho_f w^2 a_{\max}^3\,(\hat{\mathbf{w}}\cdot\p)(\hat{\mathbf{w}}\times\p)$, with $a_{\max}=\max(a,c)$ \citep{Dabade2015,Jiang2021}, and $C_r$ is the couple resisting rotation about a transverse axis,
\begin{equation}
C_r=\frac{16\pi\mu_f\,(a^2+c^2)}{3\,(a^2J_\perp+c^2J_\parallel)} .
\label{eq:Cr}
\end{equation}
We take $F_D$ from the closed-form expressions of \cite{Dabade2015} (their equations 4.1 and 4.2), with eccentricity $e=(1-\kappa^{-2})^{1/2}$ for prolate and $(1-\kappa^2)^{1/2}$ for oblate shapes; these return the near-sphere limit $811\pi e^2/1120$, the slender-body limit, and the flat-disc value $38/9-17\,216/(945\pi^2)\simeq2.376$ given there, and the prolate amplitude peaks at $\kappa\simeq1.9$. In the broadside convention of Eq.~\eqref{eq:conventions} they give $F_b=0.116$, $0.158$, $0.175$, $0.183$ and $0.189$ for columns of $\kappa=2$, $3$, $4$, $5$ and $10$, and $0.329$, $0.325$, $0.311$ and $0.320$ for plates of $\kappa=0.01$, $0.02$, $0.05$ and $0.03$; in the equivalent-sphere convention the $\kappa=10$ column has $F_{\rm eq}=0.060$. These are creeping-flow values. At the particle Reynolds numbers of real crystals, $\Rep=O(1$--$10)$, the amplitude falls below them \citep{Jiang2021}, and \cite{Pumir2026} use a calibrated reduction $C_\tau=0.5$; for slender fibres the inertial slender-body theory of \cite{Joshi2026} should be used instead.

\section{Validation summary}
\label{app:validation}

The closed-form results have been checked against quadrature or simulation as follows: the forcing law \eqref{eq:fvar} by Monte Carlo sampling ($4\times10^5$ realisations) of Gaussian $\E$ and $\W$ at four values of $\mathcal{B}$, including the tangent-plane isotropy of the strain contribution; the moments \eqref{eq:mu2pro}--\eqref{eq:absmuobl} of the Boltzmann density and the shape factor \eqref{eq:Sclosed} against $40$-digit quadrature for $b$ from $10^{-3}$ to $10^6$ on both branches, to about $10^{-12}$, the quadrature of the text using a $400$-point Gauss--Legendre rule in $\mu$ with the exponent shifted by its maximum for $b<50$, and for $b\ge50$ a $200$-point rule in the scaled variables $s=\sqrt b\,\mu$ (girdle) or $v=b(1-\mu^2)$ (bipolar), which resolve the boundary layers of width $O(b^{-1/2})$ and $O(b^{-1})$; the aligned mixture moments \eqref{eq:ssmoments}--\eqref{eq:flatness2} by quadrature at $b_0=400$ and $\sigma^2_{\ln\bar\phi}\le1.2$, to $0.01\%$ in the variance and $0.3\%$ in the flatness for a girdle and to $0.5\%$ and $2.4\%$ for a bipolar state, whose corrections are of relative order $e^{\sigma^2_{\ln\bar\phi}}/2b_0$; the random limits \eqref{eq:randpro}, the slip identities $2/15$ and $4/45$ and the relaxation \eqref{eq:tanphi} to quadrature and integrator precision; the mobility values against the sphere limit, the thin-disc speed ratio $3/2$, approached as $3/2-O(\kappa)$, and the logarithmically slow approach of $\hat U_\parallel/\hat U_\perp$ to $2$ ($1.526$ at $\kappa=20$); the identity \eqref{eq:identity} to a residual of $10^{-16}$; and the linear response \eqref{eq:linresp} by Ornstein--Uhlenbeck-forced simulation at three $(\lambda_r,\tau_c)$ pairs, to $0.3\%$.

\section{Correlation time along a settling path}
\label{app:tauc}

In the rapid-crossing, frozen-field approximation the time-integrated gradient correlation along a straight settling path follows from the energy spectrum. With $A_{ij}=\partial u_i/\partial x_j$ as in Eq.~\eqref{eq:Acov} and $\mathbf x(t')=\mathbf x(t)+\Sv\,\ghat\,(t'-t)$ in Kolmogorov units, the time integral collapses onto the plane $\mathbf k\cdot\ghat=0$,
\begin{equation}
\int_{-\infty}^{0}\!\!\langle A_{ij}(0)A_{kl}(t')\rangle\,\dd t'
=\frac{1}{4\Sv}\Bigl[\int_0^\infty\!\!k\,E(k)\,\dd k\Bigr]
\int_0^{2\pi}\!\!(\delta_{ik}-e_ie_k)e_je_l\,\dd\phi ,
\label{eq:pitensor}
\end{equation}
with $k$ the wavenumber magnitude and $\mathbf e$ a unit vector in that plane. Dividing the contracted integral by the zero-lag variance gives the correlation time of the tangential forcing, $\tau_c=(\ell/\eta)\,\tau_K/\Sv$, with $\ell/\eta$ fixed by the spectrum rather than assumed. Evaluating Eq.~\eqref{eq:pitensor} with the model spectrum of \cite{Pope2000}, at its high-Reynolds-number constants, and contracting it with Eq.~\eqref{eq:fdef} and the tangent projector at the attracting orientation, which averages over the two tangent directions, gives $\ell/\eta=12.2$, $16.4$ and $17.7$ for the design column at $\mathrm{Re}_\lambda=10^2$, $10^3$ and $10^4$, and $13.9$, $18.7$ and $20.2$ for the design plate; the integral converges as $\mathrm{Re}_\lambda\to\infty$, so $\ell$ saturates near twenty Kolmogorov lengths rather than growing with the Taylor microscale. The crossing time therefore falls below $\tau_K$ only for $\Sv\gtrsim12$--$20$ (Sec.~\ref{sec:forcing}), and where it does, as in the faster runs of table~\ref{tab:dnsfloor}, the alignment is strong enough, $\lambda_r\tau_c\gg1$, for $\tau_c$ to cancel from Eq.~\eqref{eq:b}.

The same integral shows that sampling along a settling path is axisymmetric about gravity rather than isotropic. At the girdle the meridional part of the integrated forcing exceeds the azimuthal part by the factor $(1+\mathcal{B})^2/(2+\mathcal{B}^2)$, independent of $\mathrm{Re}_\lambda$ but strongly shape dependent: $1.32$ for the $\kappa=10$ column, $1.08$ at $\kappa=2$, and $1/2$ as $\mathcal{B}\to0$, so the sense of the anisotropy reverses for nearly spherical columns. The splitting matters only in the white-noise branch (Sec.~\ref{sec:aniso}), where the tilt of a column is set by the restored, meridional component alone, whose correlation length is $2(1+\mathcal{B})^2\ell/(3+2\mathcal{B}+2\mathcal{B}^2)$, $14\%$ above the average $\ell$ for the $\kappa=10$ column. At a pole the two tangent directions are equivalent by symmetry, with no splitting.

\section{Moments of the Boltzmann density and the slip shape factor}
\label{app:moments}

Integrating $\dd(\mu^{2n-1}e^{\mp b\mu^2})/\dd\mu$ over $[0,1]$ relates successive even moments $M_n\equiv\langle\mu^{2n}\rangle$, with $M_0=1$:
\begin{equation}
M_n^-=\frac{(2n-1)M_{n-1}^--e^{-b}/Z_-}{2b},
\qquad
M_n^+=\frac{e^{b}/Z_+-(2n-1)M_{n-1}^+}{2b},
\label{eq:momrec}
\end{equation}
the superscript denoting the sign in the exponent of Eq.~\eqref{eq:watson}: the girdle branch for columns, the bipolar branch for plates. Angle moments such as $\langle\varphi^2\rangle$, with $\varphi=\arcsin m$ for a girdle and $\arccos m$ for a bipolar state, $m=|\mu|$, are evaluated by quadrature.

The single moments used in the text are, with $Z_-=\tfrac12\sqrt{\pi/b}\,\mathrm{erf}\sqrt b$ for the girdle branch,
\begin{align}
\langle\mu^2\rangle_{\rm pro} &= \frac{1}{2b}-\frac{e^{-b}}{\sqrt{\pi b}\;\mathrm{erf}\sqrt{b}},
&\langle|\mu|\rangle_{\rm pro} &= \frac{1-e^{-b}}{2b\,Z_-},
\label{eq:mu2pro}\\
\langle\mu^2\rangle_{\rm obl} &= -\frac{1}{2b}+\frac{1}{2\sqrt{b}\,\mathcal D(\sqrt b)},
&\langle|\mu|\rangle_{\rm obl} &= \frac{1-e^{-b}}{2\sqrt b\,\mathcal D(\sqrt b)},
\label{eq:absmuobl}
\end{align}
the bipolar pair written with the Dawson function for numerical stability. The random limits \eqref{eq:randpro} follow from $\int_0^{\pi/2}\varphi^2\cos\varphi\,\dd\varphi=\pi^2/4-2$ and $\int_0^{\pi/2}\varphi^2\sin\varphi\,\dd\varphi=\pi-2$, the difference being the measure on the sphere near a girdle and near a pole.

The slip shape factor of Eq.~\eqref{eq:vh2} is $\mathcal S=M_1-M_2$, so that
\begin{equation}
\mathcal S_{\rm pro}=\frac{2b-3+3\mathcal E}{4b^2},
\qquad
\mathcal S_{\rm obl}=\frac{3\mathcal G-3-2b}{4b^2},
\label{eq:Sclosed}
\end{equation}
with $\mathcal E=e^{-b}/Z_-$ and $\mathcal G=e^{b}/Z_+$ from Eq.~\eqref{eq:Znorm}, that is
\begin{equation}
\mathcal E=\frac{2}{\sqrt\pi}\,\frac{\sqrt b\,e^{-b}}{\mathrm{erf}\sqrt b},
\qquad
\mathcal G=\frac{\sqrt b}{\mathcal D(\sqrt b)} .
\label{eq:EG}
\end{equation}
Both reduce to $2/15$ as $b\to0$, and for strong alignment
\begin{equation}
\mathcal S_{\rm pro}\to\frac{1}{2b}-\frac{3}{4b^2},
\qquad
\mathcal S_{\rm obl}\to\frac1b-\frac{3}{2b^2} .
\label{eq:Sasym}
\end{equation}
Equation~\eqref{eq:Sclosed} loses precision below $b\simeq0.05$, where the numerator is a difference of nearly equal terms, and quadrature should be used there. The vertical fluctuation follows from the same moments,
\begin{equation}
\mathrm{var}(v_z)=\Sv^2 g_M^2\bigl(M_2-M_1^2\bigr),
\label{eq:vzvar}
\end{equation}
whose bracket tends to $4/45$ in the random limit and falls as $\Sv^{-8}$ once aligned, so that $\mathrm{var}(v_z)\propto\Sv^{-6}$: near a pole $1-\mu^2$ is exponentially distributed with variance $b^{-2}$, and near a girdle $\mu^2$ is a scaled one-degree-of-freedom chi-square variable.

The threshold quoted for the slip ratio \eqref{eq:ratio} in Sec.~\ref{sec:invert} follows from the same moments. With $t=\hat U_\parallel/\hat U_\perp-1$ and primes denoting derivatives with respect to $b$ on the girdle branch, where $M_1'=-(M_2-M_1^2)<0$,
\begin{equation}
\frac{\dd\ln\mathcal R_v}{\dd b}=\frac{\mathcal S'}{2\mathcal S}+\frac{t\,(M_2-M_1^2)}{1+tM_1},
\label{eq:Rvslope}
\end{equation}
which increases with $t$ at fixed $b$. Near random orientation $M_1=\tfrac13-\tfrac{4}{45}b$ and $\mathcal S=\tfrac{2}{15}-\tfrac{4}{315}b$, so at $b=0$ the slope is $-\tfrac{1}{21}+4t/[45(1+t/3)]$, which vanishes at $t=15/23$, that is $\hat U_\parallel/\hat U_\perp=38/23$. At that ratio Eq.~\eqref{eq:Rvslope} is negative for all $b>0$ (evaluated for $10^{-8}\le b\le10^6$), so $\mathcal R_v$ falls monotonically for every column with $\hat U_\parallel/\hat U_\perp<38/23$; for larger ratios it rises first, by an amount that grows with the ratio and tends to about $0.72\%$ in the slender limit $\hat U_\parallel/\hat U_\perp\to2$.

\section{Conditional forcing statistics}
\label{app:cond}

Resolving the forcing in the tangent frame, into the meridional direction $\hat{\mathbf e}_1\propto\ghat-\mu\p$, along which the tilt is restored, and the azimuthal direction $\hat{\mathbf e}_2=\p\times\hat{\mathbf e}_1$, gives the second moments of table~\ref{tab:cond}. They are not equal, and their ratio varies across the sphere, from $0.64$ to $2.33$ for the column and from $2.1$ to $4.3$ for the plate.

Stationarity fixes the conditional mean. Since $\dot\mu=\sqrt{1-\mu^2}\,f_1-s_\kappa\lambda_r\mu(1-\mu^2)$, a stationary marginal with no probability current requires
\begin{equation}
\langle f_1\mid\mu\rangle=s_\kappa\lambda_r\,\mu\sqrt{1-\mu^2} ,
\label{eq:condmean}
\end{equation}
which the simulations reproduce to $0.2\%$ when binned in signed $\mu$ and the right-hand side is averaged over the samples in each bin: for the $\kappa=10$ column at $\Sv=1.78$ the measured values in bins centred at $\mu=0.1$ to $0.9$ are $0.093$, $0.270$, $0.416$, $0.493$ and $0.406$, and the bin averages of Eq.~\eqref{eq:condmean} agree to the precision shown, whereas its values at the bin centres, $0.100$, $0.287$, $0.434$, $0.500$ and $0.393$, differ by up to $7\%$ through the curvature within the bins. Binning in $|\mu|$ cancels it, since the mean is odd in $\mu$. Equation~\eqref{eq:condmean} is neither a torque of turbulent origin nor a correction the closure omits (Sec.~\ref{sec:conditioning}), and its agreement with the simulations checks their stationarity rather than the closure. Because the mean is non-zero, the entries of table~\ref{tab:cond} are raw second moments and not centred variances; at $|\mu|=0.58$ the mean contributes about three quarters of $\langle f_1^2\rangle$ for the column and two thirds for the plate.

These measurements show how the sampled forcing departs from the fixed-orientation forcing, but they do not separate the drift and memory corrections to the generator: $\langle f_1^2\rangle\tau_1$, with $\tau_1$ the integral time of $f_1$, is an ordinary correlation integral, whereas $D_{\rm eff}$ is response weighted, so the coloured-noise factor need not absorb them.

\begin{table}
\caption{\label{tab:cond}Raw second moments of the forcing conditioned on orientation, from tensor-forced runs at $\Sv=1.78$ ($N=4000$, $200\tau_K$, the first $40\tau_K$ discarded; Appendix~\ref{app:numerics}), resolved into the meridional and azimuthal tangent directions and binned in $|\mu|$. Standard errors over $16$ groups of particles are below $0.002$ except in the least populated bins, where they reach $0.006$. The fixed-orientation value assumed by Eq.~\eqref{eq:fvar} is $\sigma_f^2\simeq0.131$ for the column and $0.133$ for the plate.}
\centering\small
\begin{tabular*}{\textwidth}{@{\extracolsep{\fill}}lcccccc}
\hline
& \multicolumn{3}{c}{$\kappa=10$ (attractor $\mu=0$)} & \multicolumn{3}{c}{$\kappa=0.03$ (attractor $\mu=1$)}\\
$|\mu|$ & $\langle f_1^2\rangle$ & $\langle f_2^2\rangle$ & ratio & $\langle f_1^2\rangle$ & $\langle f_2^2\rangle$ & ratio\\
\hline
$0.08$ & $0.073$ & $0.114$ & $0.64$ & $0.222$ & $0.105$ & $2.11$\\
$0.25$ & $0.123$ & $0.117$ & $1.05$ & $0.254$ & $0.103$ & $2.47$\\
$0.42$ & $0.208$ & $0.123$ & $1.70$ & $0.310$ & $0.096$ & $3.23$\\
$0.58$ & $0.300$ & $0.133$ & $2.26$ & $0.353$ & $0.088$ & $4.04$\\
$0.75$ & $0.353$ & $0.151$ & $2.33$ & $0.330$ & $0.077$ & $4.29$\\
$0.92$ & $0.312$ & $0.179$ & $1.74$ & $0.154$ & $0.068$ & $2.27$\\
\hline
\end{tabular*}
\end{table}

\section{Numerical methods}
\label{app:numerics}

The Langevin simulations integrate Eq.~\eqref{eq:langevin} ($-$ prolate, $+$ oblate) with two forcings. The projected-vector forcing is $\mathbf f=(\I-\p\p)\cdot\bm{\xi}$, with $\bm{\xi}$ an isotropic Ornstein--Uhlenbeck vector of per-component variance $\sigma_f^2$ and correlation time $\tau_c$; it has the covariance \eqref{eq:fvar} at a fixed orientation but not the dynamics of the tensor, which enters through the orientation-dependent map \eqref{eq:fdef}. The tensor forcing evolves $\E$ and $\W$ themselves. At each step a Gaussian matrix $\mathsf G$ with independent standard normal entries is split into its traceless symmetric and antisymmetric parts,
\begin{equation}
\mathsf S=\tfrac12(\mathsf G+\mathsf G^{\!\top})-\tfrac13(\operatorname{tr}\mathsf G)\I,
\qquad
\mathsf A=\tfrac12(\mathsf G'-\mathsf G'^{\!\top}),
\label{eq:tensdraw}
\end{equation}
with $\mathsf G'$ an independent draw, and $\E$ and $\W$ are advanced by the exact update
\begin{equation}
\E_{k+1}=e^{-\Delta t/\tau_c}\E_k+a_E\sqrt{1-e^{-2\Delta t/\tau_c}}\;\mathsf S_k,
\qquad
\W_{k+1}=e^{-\Delta t/\tau_c}\W_k+a_W\sqrt{1-e^{-2\Delta t/\tau_c}}\;\mathsf A_k .
\label{eq:tensou}
\end{equation}
Since $\langle\mathsf S_{12}^2\rangle=\langle\mathsf A_{12}^2\rangle=\tfrac12$ for these draws, $a_E^2=1/10$ and $a_W^2=1/6$ in Kolmogorov units reproduce $\langle E_{12}^2\rangle=\varepsilon/20\nu$ and $\langle W_{12}^2\rangle=\varepsilon/12\nu$ from Eqs.~\eqref{eq:Ecov}--\eqref{eq:Wcov}. The realised statistics reproduce $\langle E_{12}^2\rangle$, $\langle E_{11}^2\rangle$, $\langle W_{12}^2\rangle$ and both traces to better than one per cent, with $\operatorname{tr}\E$ zero to machine precision. Halving the time step to $0.005\tau_K$ changes $\langle\mu^2\rangle$ by less than $1\%$, within about two standard errors, over $\lambda_r\tau_K=0.45$--$2.2$, and five independent runs with each forcing at $\Sv=1.78$, with $N=2500$ and the first $40\tau_K$ discarded, over $110\tau_K$ (projected vector) and $170\tau_K$ (tensor), scatter by $0.4\%$ and $0.5\%$ about their means, so the discrepancies reported in Sec.~\ref{sec:watson} are not numerical. The densities of figure~\ref{fig:sde}(b,c) come from tensor runs with $N=3500$ particles over $170\tau_K$, the first $40\tau_K$ discarded, binned in $40$ equal intervals of $|\mu|$, each bin value being plotted where the closure density equals its own average over the bin. The vector is updated exactly over each step, $\bm{\xi}_{k+1}=e^{-\Delta t/\tau_c}\bm{\xi}_k+\sigma_f\sqrt{1-e^{-2\Delta t/\tau_c}}\,\mathbf r_k$ with $\mathbf r_k$ independent standard normal vectors, and $\p$ is advanced by a Heun predictor--corrector with renormalisation. At finite $\tau_c$ the forcing is continuous in time, though not differentiable, so the orientation equation is a random ordinary differential equation and no It\^o correction arises.

Figure~\ref{fig:sde}(a) shows the projected-vector runs of the next paragraph, at $\Sv=(\lambda_r\tau_K/F)^{1/2}$ with $F=0.316$, and figure~\ref{fig:drift}(b) the slip from the same runs. For the amplitude-modulated runs of Sec.~\ref{sec:intermittency}: the $\kappa=10$ column with $\tau_c=\tau_K$ and $\lambda_r\tau_K=3$, forced by $\mathbf f=\bar\phi^{1/2}(t)\,(\I-\p\p)\cdot\bm\xi(t)$, with $\bm\xi$ the unmodulated vector above and $\ln\bar\phi$ an independent Ornstein--Uhlenbeck process for each particle, of mean $-\tfrac12\sigma^2_{\ln\bar\phi}$, variance $\sigma^2_{\ln\bar\phi}=1.2$ and correlation time $\tau_{\bar\phi}$, updated exactly; $\Delta t=0.01\tau_K$; $N=4000$ particles over $100$, $100$ and $200\tau_K$ for $\tau_{\bar\phi}/\tau_K=0.3$, $1$ and $4$, and $N=3000$ over $600\tau_K$ for $\tau_{\bar\phi}=20\tau_K$, discarding the first $20$, $20$, $30$ and $100\tau_K$, with two to four independent runs per case. The frozen-amplitude value is a $16$-point Gauss--Hermite quadrature in $\ln\bar\phi$ over runs at fixed amplitude ($N=3000$, $60\tau_K$, the first $20\tau_K$ discarded), and the reference is $\int P(\bar\phi)\,\langle\mu^2\rangle_{b_c/\bar\phi}\,\dd\bar\phi$, with $b_c$ the closure concentration \eqref{eq:b} at uniform dissipation. The fits to the data of \cite{Anand2020} minimise the unweighted rms residual of $\ln P$ over the digitised points, every density being normalised on $0\le m\le1$ and the mixture \eqref{eq:superstat} evaluated by $80$-point Gauss--Hermite quadrature in $\ln\bar\phi$; the four points of figure~\ref{fig:anand}(b) are the centroids of the symbols in the embedded raster of their Fig.~2(b), calibrated against the frame of the inset. The moments of figure~\ref{fig:anand}(c,d) come from the main panels of the same raster: each logarithmic axis is calibrated on its decade ticks, the filled triangles of the $\mathrm{Re}_\lambda=150$ runs are segmented by colour, and each is placed at the centre of its bounding box, where one pixel is $2.1\%$ in $X$ and $3.9\%$ in $Y$; two markers near $\mathrm{Fr}_\eta=2.7$ overlap and are not used. Each marker is matched to a run of their supplementary table by the nearest tabulated $\mathrm{Fr}_\eta$, and $\Sv_b=\mathrm{Fr}_\eta$ is read from the plotted abscissa. Placing the markers at their pixel centroids instead, or taking $\mathrm{Fr}_\eta$ from the rounded values of the table, keeps the single density within $10\%$ and the mixture within $15\%$ of the measured $X$ from $F_b\Sv_b^2=0.8$ upwards, gives factors of $2.5$--$3.1$ and excesses of $9$--$54\%$ in $Y$ from $F_b\Sv_b^2=1.3$ upwards, and a weak-torque departure ratio of $1.3$--$2.2$. For the data of \cite{Pumir2026} we use the vertices of the curves of their Fig.~2, read from the vector paths of the published article and converted with its axis frame; the vertices sit at the centres of histogram bins of width $0.01$ and are quantised in steps of $1.5\times10^{-4}$ in density, so points below $1.5\times10^{-3}$ are not used, and each fit minimises the unweighted squared deviation of $A\exp(-bm^2)$ from the vertices above a fiftieth of the peak. Table~\ref{tab:dnsfloor} uses the parameters of their Tables~I and II, $\nu=0.1875\,\mathrm{cm^2\,s^{-1}}$, $\rho_f=9.022\times10^{-4}\,\mathrm{g\,cm^{-3}}$ and $g=981\,\mathrm{cm\,s^{-2}}$, with buoyancy neglected as in their response time, and the drag estimate of Sec.~\ref{sec:application} uses their equations (C3) and (C4) with their drag correction $C_f=0.6$.

The second moments quoted in Sec.~\ref{sec:watson} come from repeat runs with both forcings, $N=4000$, $\Delta t=0.01\tau_K$ and total time $200\tau_K$, the first $40\tau_K$ discarded, with standard errors of $0.1$--$0.4\%$. For the slip maximum of Sec.~\ref{sec:peak}: tensor forcing at $\lambda_r\tau_K=e^{x}$, $x$ from $-0.4$ to $0.2$ (column) or $-0.2$ to $0.4$ (plate) in steps of $0.1$, all runs of a shape driven by the same random sequence; $N=8000$ in $16$ groups, the same timing, at $\Delta t=0.01\tau_K$ and $0.02\tau_K$. The maximum of $u\mathcal S$ is located from a cubic spline through $\ln(\mathcal S/\mathcal S_{\rm cl})$ added to the closure curve \eqref{eq:Sclosed}, and checked against a cubic fit of $\ln u\mathcal S$, a local quadratic, and the crossing of $-1$ by the finite-difference slope of $\ln\mathcal S$; the sampling error is one standard deviation over $400$ bootstrap resamplings of the groups, the same at every $u$. Independent runs on the coarser grid $\lambda_r\tau_K=0.45$--$2.2$, at $\Delta t=0.01\tau_K$ and $0.005\tau_K$, agree within their larger sampling error. For the azimuthal test of Sec.~\ref{sec:watson}: the column under the projected-vector forcing at $\lambda_r\tau_K=3$ and $6$ ($N=4000$, $120\tau_K$, the first $30\tau_K$ discarded, eight seeds for each case, the two cases sharing their random numbers seed by seed), with and without the azimuthal component of $\mathbf f$: $\langle\mu^2\rangle$ is $0.956$ and $0.972$ of the linear response \eqref{eq:linresp} with it and $1.027$ and $1.008$ without, each with a standard error of $0.0015$, and the mean of $\bm\xi$ along the horizontal direction of $\p$, $0.118$ with it and zero without, accounts as an added restoring rate for over $90\%$ of the difference. The forcing statistics of Sec.~\ref{sec:conditioning} come from projected-vector runs of the column with the same timing at $\lambda_r\tau_K=1.0$ and $0.079$, two seeds each.

Moments of the Boltzmann density are evaluated as in Appendix~\ref{app:validation}, and the lateral-slip maxima of Eq.~\eqref{eq:ustar} were located by bounded scalar minimisation to $10^{-5}$.

The digitisation of the published data and the simulations of this appendix were carried out between August and November 2026 with Claude Opus 5 to 5.5 (Anthropic), which wrote and ran the extraction scripts and the simulation drivers; their availability is given in the data availability statement.

\bibliographystyle{jfm}
\bibliography{tiltwatson}

\end{document}